\documentclass[preprint]{aastex}
\usepackage{lineno}
\usepackage{natbib}
\usepackage{amssymb}
\usepackage{float}
\usepackage{lscape}
\shorttitle{The FAVA source catalog}
\shortauthors{Author et al.}

\begin{document}

\title{The \emph{Fermi} All-sky Variability Analysis: \\
A list of flaring gamma-ray sources and the search for transients in our Galaxy}
\author{
M.~Ackermann\altaffilmark{1}, 
M.~Ajello\altaffilmark{2,3}, 
A.~Albert\altaffilmark{4}, 
A.~Allafort\altaffilmark{5,6}, 
E.~Antolini\altaffilmark{7,8}, 
L.~Baldini\altaffilmark{9}, 
J.~Ballet\altaffilmark{10}, 
G.~Barbiellini\altaffilmark{11,12}, 
D.~Bastieri\altaffilmark{13,14}, 
K.~Bechtol\altaffilmark{5}
% ...
}
\altaffiltext{1}{Deutsches Elektronen Synchrotron DESY, D-15738 Zeuthen, Germany}
\altaffiltext{2}{Space Sciences Laboratory, 7 Gauss Way, University of California, Berkeley, CA 94720-7450, USA, , USA}
\altaffiltext{3}{email: majello@slac.stanford.edu}
\altaffiltext{4}{Department of Physics, Center for Cosmology and Astro-Particle Physics, The Ohio State University, Columbus, OH 43210, USA}
\altaffiltext{5}{W. W. Hansen Experimental Physics Laboratory, Kavli Institute for Particle Astrophysics and Cosmology, Department of Physics and SLAC National Accelerator Laboratory, Stanford University, Stanford, CA 94305, USA}
\altaffiltext{6}{email: allafort@stanford.edu}
\altaffiltext{7}{Dipartimento di Fisica, Universit\`a degli Studi di Perugia, I-06123 Perugia, Italy}
\altaffiltext{8}{Istituto Nazionale di Fisica Nucleare, Sezione di Perugia, I-06123 Perugia, Italy}
\altaffiltext{9}{Universit\`a  di Pisa and Istituto Nazionale di Fisica Nucleare, Sezione di Pisa I-56127 Pisa, Italy}
\altaffiltext{10}{Laboratoire AIM, CEA-IRFU/CNRS/Universit\'e Paris Diderot, Service d'Astrophysique, CEA Saclay, 91191 Gif sur Yvette, France}
\altaffiltext{11}{Istituto Nazionale di Fisica Nucleare, Sezione di Trieste, I-34127 Trieste, Italy}
\altaffiltext{12}{Dipartimento di Fisica, Universit\`a di Trieste, I-34127 Trieste, Italy}
\altaffiltext{13}{Istituto Nazionale di Fisica Nucleare, Sezione di Padova, I-35131 Padova, Italy}
\altaffiltext{14}{Dipartimento di Fisica e Astronomia "G. Galilei", Universit\`a di Padova, I-35131 Padova, Italy}
% ...


\begin{abstract}
In this paper we present the \emph{Fermi} All-sky Variability Analysis (FAVA), a tool to systematically study the variability of the gamma-ray sky measured by the Large Area Telescope (LAT) on board the \emph{Fermi} Gamma-ray Space Telescope. For each direction on the sky, FAVA compares the number of gamma rays observed in a given time window to the number of gamma rays expected for the average emission detected from that direction. This method is used in weekly time intervals to derive a list of 215 flaring gamma-ray sources. We proceed to discuss the 27 sources found at Galactic latitudes smaller than 10$^{\circ}$ and show that, despite their low latitudes, most of them are likely of extragalactic origin.
\end{abstract}


\maketitle 

\section{Introduction}

In 1844 the astronomer F. W. A. Argelander performed one of the first systematic studies of the variability of the night sky. He laid the study of variable sources ``most pressingly on the heart of all lovers of the starry heavens, to perform an important part towards the increase of human knowledge, and help to investigate the eternal laws which announce in endless distance the Almighty power and wisdom of the Creator'' \citep{Percy2007}. Nowadays astronomers are not as poetic, but time has provided us with exceptional instruments for the quest.

In this paper we present a systematic study of the temporal variations of the gamma-ray sky measured by the LAT on board the \emph{Fermi}  Satellite. The gamma-ray sky above 100 MeV is dominated by the Galactic diffuse emission, which originates from cosmic-ray interactions with interstellar matter and photon fields (e.g. \citealt{Ackermann2012a}). Additionally, an isotropic diffuse gamma-ray emission is detected, and it is the strongest source of emission at high Galactic latitudes \citep{Abdo2010c}. Both diffuse components are expected to be stable over the duration of the \emph{Fermi} mission. On-top of this background, 1873 gamma-ray sources have been detected during the first two years of Fermi mission and reported in the second \emph{Fermi}-LAT catalog (2FGL, \citealt{Nolan2012}). Out of these sources $\sim$24\% are found to be variable on monthly time scales. The vast majority of the variable sources are associated with Active Galactic Nuclei (AGN), which are known to be variable across the electromagnetic spectrum (e.g. \citealt{Aharonian2009,Ackermann2011a,Ackermann2011b}). 

Variability in gamma rays has so far been established only for a few sources in our Galaxy. Orbital modulation and isolated flares have been reported from 7 X-ray binaries, in which a neutron star or a black hole orbits a massive companion \citep{Aharonian2005,Aharonian2006,Albert2006,Abdo2009,Hinton2009,Sabatini2010,FermiLATCollaboration2012}. Variable gamma-ray emission has also been reported from the direction of Eta Car, a massive star-Wolf-Rayet binary \citep{Tavani2009,Abdo2010a,Reitberger2012}.  Recently, flaring gamma-ray emission has been found for two new source classes: Nova explosions \citep{Abdo2010} and the Crab pulsar wind nebula \citep{Abdo2011,Tavani2011,Buehler2012}. The latter was thought to be a stable gamma-ray emitter, but was discovered to be flaring with the method described in this paper. In addition several other gamma-ray transients have been detected near  the Galactic equator, but they are likely associated with distant AGN \citep{Vandenbroucke2010,Cheung2012}.

The reasons for the detection of so few variable gamma-ray sources within our Galaxy remain unclear, whether due to astrophysical reasons, due to the statistically limited flux sensitivity, or due to systematic difficulties of detecting them owing to uncertainties in the modeling of the strong foreground of the Galactic diffuse emission.  In this paper we present a new method developed to search for transients in the gamma-ray sky that does not require a diffuse emission model. We first describe the methods and proceed to assemble a list of flaring gamma-ray sources  seen over the sky during the first 47 months of the \emph{Fermi} mission. We then focus on the sources detected at low Galactic latitudes, as they may be of Galactic origin.

\section{The \emph{Fermi} All-sky Variability Analysis}

\begin{figure*}[ht]
\begin{center}
\begin{tabular}{cl}
 \includegraphics[width=0.9\textwidth]{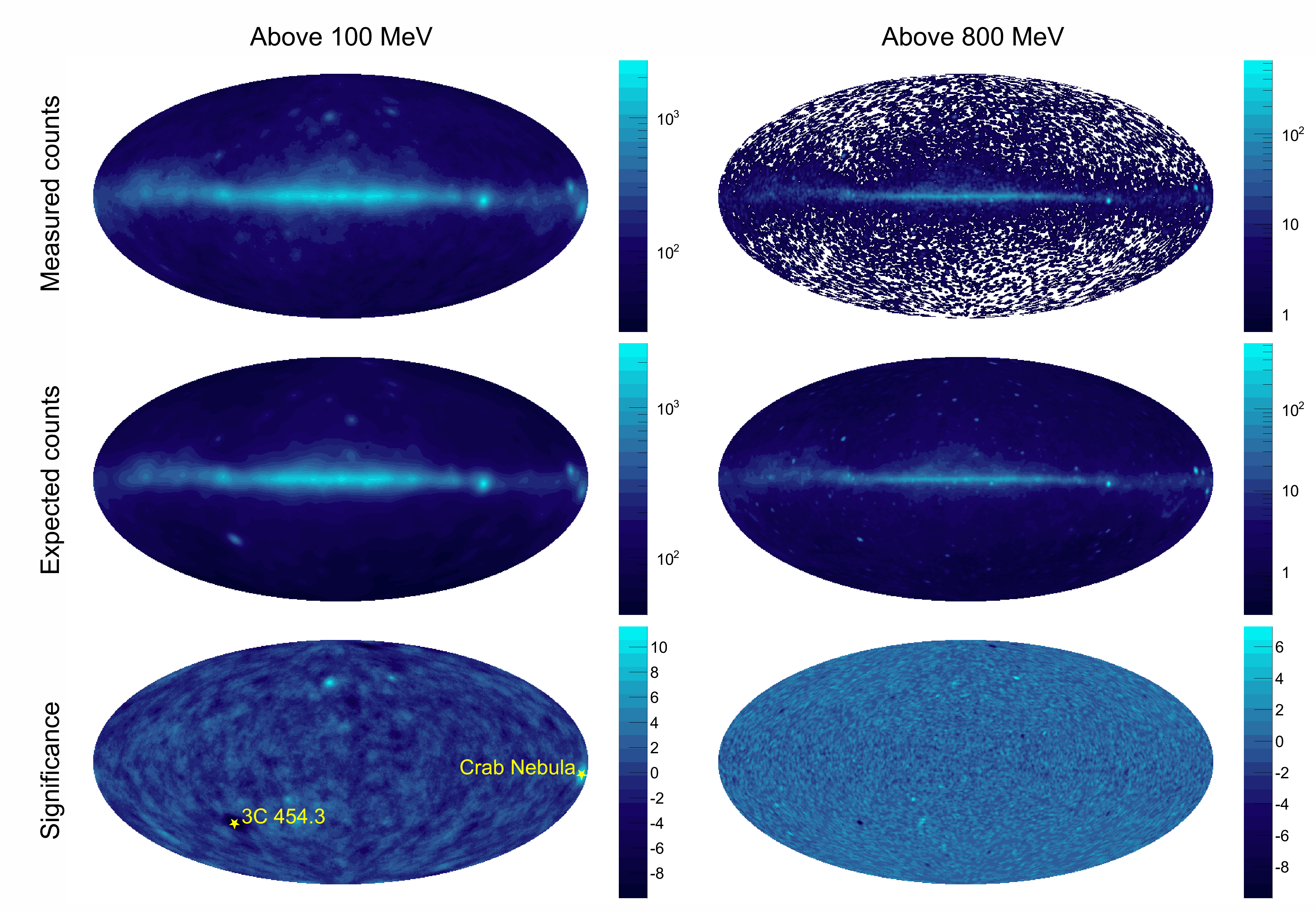} &
\end{tabular}
\end{center}
\caption{Illustration of the map generation in FAVA. The week shown covers the time interval MJD 54864.655---54871.655 (2009-02-02 to 2009-02-09). The measured counts map is shown in the first row, for energies $>100$ MeV (left) and energies $>800$ MeV (right, white color indicates that no counts were detected from this area).  The middle panels show the expected counts from the average emission observed during the first 47 months of \emph{Fermi} observations. The generation of measured and expected counts is explained in the text. The third row shows the significance of the flux variations. As an example the position of the Crab Nebula is indicated by a star in the lower left panel. Its flux was increased compared to average with a significance $>5\sigma$ for energies $>100$ MeV during this week. The flare is only detected in the low energy range, as the energy spectrum was very soft (photon index $\sim$3.5, see \citealt{Abdo2011}). An example of a flare detection with a negative flux variation is given by the blazar 3C 454.3. Its flux was lower than average  for both the $>$100 MeV and $>$800 MeV energy ranges with a significance $>8\sigma$. Figures are shown in Galactic coordinates in a Hammer-Aitoff projection. Note that the color scales were adjusted to different ranges for the low- and high-energy bands.}
\label{fig:maps}
\end{figure*}

Flux variability of LAT sources is usually studied with a maximum likelihood analysis, in which parameters of a model describing the point sources and diffuse gamma-ray emission in a given region of the sky are jointly optimized. The sensitivity of this approach is often limited by the uncertainties of the diffuse emission modeling, particularly in the Galactic plane  \citep{Nolan2012}. Small inaccuracies in the instrument response functions can lead to time-dependent residuals which  depend on varying observation conditions, e.g. off-axis angle of the sources or orbital position of the LAT, limiting variability studies \citep{Ackermann2012}. A further limiting factor of the likelihood approach is that it is computationally intensive; it is currently difficult to perform variability studies in different time and energy windows over the entire sky. We therefore developed  the \emph{Fermi} All-sky Variability Analysis (FAVA), in which we search over a grid of regions on the sky for deviations from the expected flux based on the long-term average. While this approach is less sensitive than a likelihood analysis, it has three main advantages:
\begin{enumerate}
\item The analysis is independent of any model for the diffuse gamma-ray emission. The diffuse emission is expected to be constant over the time of the \emph{Fermi} mission. It therefore cancels out in the comparisons between number of expected and measured gamma-ray events. 
\item The analysis is computationally inexpensive, allowing us to blindly search for flux variations over the entire sky. The analysis is therefore unbiased, treating every direction on the sky equally, potentially yielding unexpected discoveries.
\item No assumptions are made about the spectral shapes of the gamma-ray sources. Negative flux variations are treated the same way as positive ones. (Throughout this paper we refer to both positive and negative variations from the mean as flares.)
\end{enumerate}


We applied FAVA to the first 47 months of \emph{Fermi} observations	(2008-08-04 to 2012-07-16 UTC), in weekly time intervals. The total number of weeks is 206. We considered two ranges of gamma-ray energy, $E >$ 100 MeV and $E >$ 800 MeV, to increase the sensitivity for spectrally soft and hard flares, respectively. We used the P7SOURCE\_V6 event selection and only considered events with a zenith angle smaller than 95$^\circ$, to limit contamination from the gamma-ray emission of the Earth Atmosphere, which is time variable in sky coordinates.

We generate measured and expected counts maps with a resolution of $0.25$ square degrees per pixel. The maps are smoothed by assigning to each pixel all events that were detected within a distance corresponding to the 68\% containment radius of the Point Spread Function (PSF). The pixel positions are characterized in spherical coordinates by $\phi$ and $\theta$. The number $N^{exp}(\phi,\theta)$ of expected events in one pixel is derived from the number $N^{tot}(\phi,\theta)$ of events observed from the same direction over the first 47 months of observations. As the PSF depends on the photon energy $E$ and on the incidence angle $\alpha$ with respect to the LAT \citep{Ackermann2012}, we integrate over these parameters:
\begin{equation}
N^{exp}(\phi,\theta)=\sum_{E: j=1..12}\sum_{\alpha: i=1..4} N^{tot}_{i,j}(\phi,\theta) \times \frac{\epsilon_{i,j}^{week}(\phi,\theta)}{\epsilon_{i,j}^{tot}(\phi,\theta)},
\end{equation}
where $\epsilon^{week}$ and $\epsilon^{tot}$ are weekly and total exposures. We proceed to calculate the probability that the observed counts are a statistical fluctuation of the expected value, based on Poisson statistics. These probabilities are then converted to significance in units of Gaussian sigmas for easier visualization. The procedure is illustrated in Figure \ref{fig:maps} for one example week in 2009 February, during which the first flare from the Crab Nebula was seen in LAT data. 

\begin{figure*}[h]
\begin{center}
\begin{tabular}{cl}
\hspace{-1.8cm}
 \includegraphics[width=1.1\textwidth]{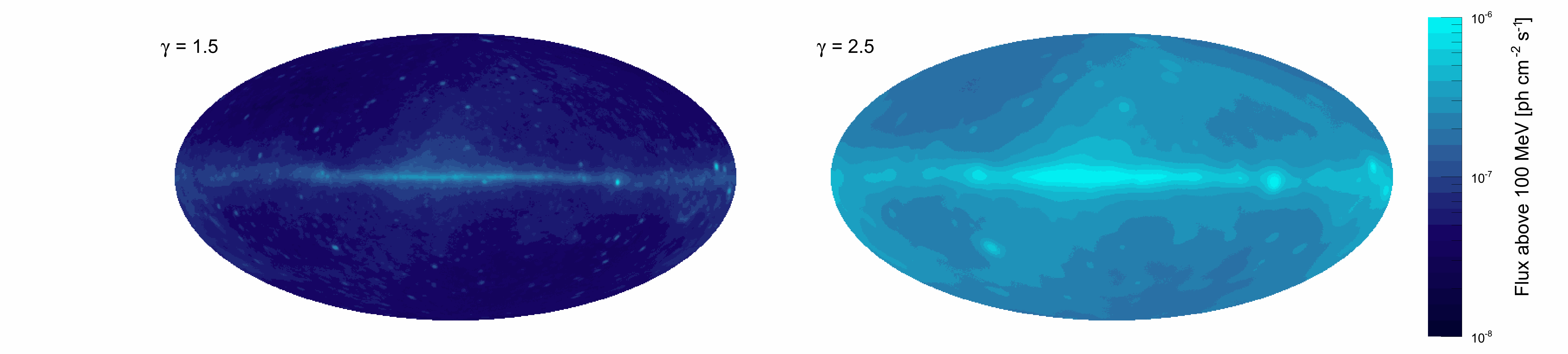} &
\end{tabular}
\end{center}
\caption{Minimum flux increase required for a flare detection in FAVA in a one-week interval  with a significance of 5.5$\sigma$ in the low-energy or the high-energy band. Detection above these thresholds results in inclusion in the list of flaring sources, as discussed in section \ref{sec:cat}. The figures are shown in Galactic coordinates in a Hammer-Aitoff projection. The flaring source is assumed to have a power-law spectrum in energy. The left panel shows the sensitivity for spectrally hard flares (photon index 1.5), the right panel for soft flares (photon index 2.5).}
\label{fig:sens}
\end{figure*}

The sensitivity of FAVA to detect flares varies with the position in the sky, due to the anisotropy of the diffuse emission backgrounds. Additionally, the sensitivity depends on the energy spectrum of the flaring gamma-ray source. Typically gamma-ray sources have photon indices between 1.5 and 2.5 in the \emph{Fermi}-LAT energy range \citep{Nolan2012}. The sensitivity for both of these cases are shown in Figure \ref{fig:sens}. Flares with a photon index greater than $\sim$2 are typically detected at higher significance in the low-energy maps, while those with a smaller index are detected more significantly in the high-energy maps.

\begin{figure*}[ht]
\begin{center}
\begin{tabular}{cl}
 \includegraphics[width=0.9\textwidth]{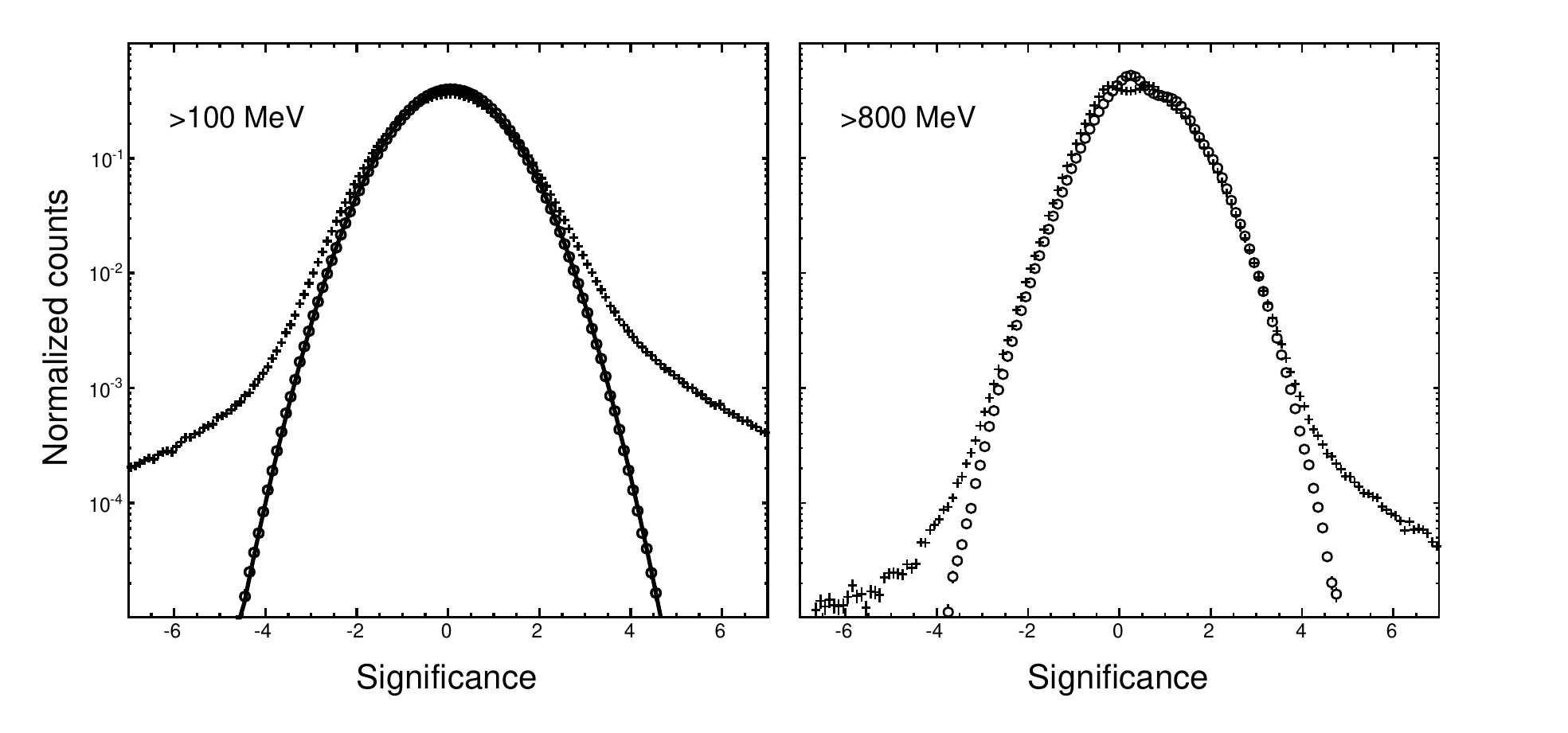} &
\end{tabular}
\end{center}
\caption{Distribution of the significance of flux variations for the low- (left) and high- energy (right) intervals. The integral of both distributions has been normalized to one. Open circles show the simulations of a constant gamma-ray sky. Crosses show the measurement of the first 47 months of \emph{Fermi} observations. The solid line in the left panel shows the best-fit Gaussian model for the simulated distribution (see text). }
\label{fig:sigdist}
\end{figure*}

The accuracy of FAVA was tested on simulations of a constant sky for 36 months of observations. The simulated sky was composed of the Galactic (gal\_2yearp7v6\_v0.fits) and isotropic (iso\_p7v6source.txt) diffuse emission models\footnote{http://fermi.gsfc.nasa.gov/ssc/data/access/lat/BackgroundModels.html.} and point sources. The latter were generated with random coordinates on the sky and according to the flux distribution derived in \citet{Abdo2010b}. The distribution of the significance $\sigma$ of flux variations is displayed in Figure \ref{fig:sigdist} for the simulations and the real data. In the low-energy interval, on average 154 events are recorded per sky pixel in each week. The simulated significance distribution is therefore expected to be close to Gaussian in the low-energy band. Indeed, a fit to the simulations shows that a Gaussian model with a standard deviation of $1.00$ and a mean of $0.06$ describes the distribution well, as shown in Figure \ref{fig:sigdist}. In the high-energy band an average of 3.6 events are recorded per sky pixel. The simulated distribution of flux variations is therefore not expected to be Gaussian. In particular, typically only a few counts are detected in several pixels at higher Galactic latitudes, leading to small-scale structure around $|\sigma|<1$ in the significance distribution. A disagreement is visible in this region between the data and simulations. However, for $|\sigma|>2$ the qualitative agreement with expectations is good also in the high-energy band, we therefore expect no significant biases in the detection of flares at high significance. As at low energies, one can see that the difference between simulated and real significance distributions increases with increasing $\sigma$, corresponding to the real flux variations emerging above the statistical background.  

The good agreement of the statistical fluctuations in the low-energy band of the simulated constant sky with expectations shows that systematic effects are small in the simulations. However, as seen for the discrepancy in the $|\sigma|<1$ interval in the high-energy band, additional effects might be present in the real data, as the simulations do not take into account possible sources of systematic errors. In particular, they do not account for any disagreement between the simulated and real instrument responses, nor for varying background levels due to residual cosmic rays mistakenly classified as gamma rays. It is difficult to assess these systematic effects from the data, as e.g. small flux variations might be present throughout the gamma-ray sky due to variable background sources. However, we can set upper limits on possible systematic errors by looking at presumably constant sources such as pulsars. Analysis of the brightest ones, the Vela and Geminga pulsars, shows that their relative count variations are compatible with a steady flux within $<$5\% on weekly time scales. Systematic errors of FAVA for relative flux variations of bright sources are therefore comparable to those for the standard \emph{Fermi}-LAT analysis \citep{Ackermann2012}.

\section{List of flaring gamma-ray sources}
\label{sec:cat}
After calculating the significance maps for all weeks, we scan them for significant flares. To determine their positions we use the peak finding algorithm described in \citet{Morhac2000}.  Based on these detections, we build a list of flaring sources. For this we only consider flares with significances greater than 5.5 $\sigma$ in the low-energy  or high-energy band.  This threshold was chosen so that the number of false flare detections due to statistical fluctuations is expected to be $\sim1$ over the 206 weeks that were analyzed\footnote{The number of trials can be approximately estimated as the total sky area divided by the area of the PSF. Above 100 MeV the 68\% containment radius of the event-averaged PSF of the \emph{Fermi}-LAT is $\sim 3^{\circ}$. We therefore have $\sim 41253/(\pi \times 3^2) \times 206 =300558$ independent tests in the sky. This results in $\sim$0.01 expected false positives above 5.5$\sigma$. For the high-energy maps the average PSF is $\sim 0.6^{\circ}$, resulting in $\sim$0.3 expected false positives above the same threshold.}. No flares were detected in the simulations of three years of data for a steady sky above this threshold. The number of false flares expected for 47 months of data is therefore $<$3 at 90\% confidence. Additionally, we only consider flares that occurred far away from the average position of the Sun in the corresponding week. The Sun is a bright gamma-ray source that moves along the ecliptic by $\sim7^\circ$ per week \citep{Abdo2011a}. We therefore only considered flares at a distance from the Sun $>$12$^\circ$ and $>$8$^\circ$ in the low- and high-energy bands, respectively. Finally, we merge low- and high-energy flares detected in the same week, if they are coincident in position within $3^{\circ}$, relating them to the same flaring event. For the position of the latter  we use the position of the high-energy detection, due to its higher accuracy, as will be discussed in the next paragraph. A total of 1419 flares that fulfill the mentioned criteria were detected. Out of these, 645 and 175 are detected in the low- and high-energy interval only, respectively. The remaining 599 flares are detected simultaneously in both energy bands.

\begin{figure}[h]
\begin{center}
\begin{tabular}{cl}
 \includegraphics[width=0.45\textwidth]{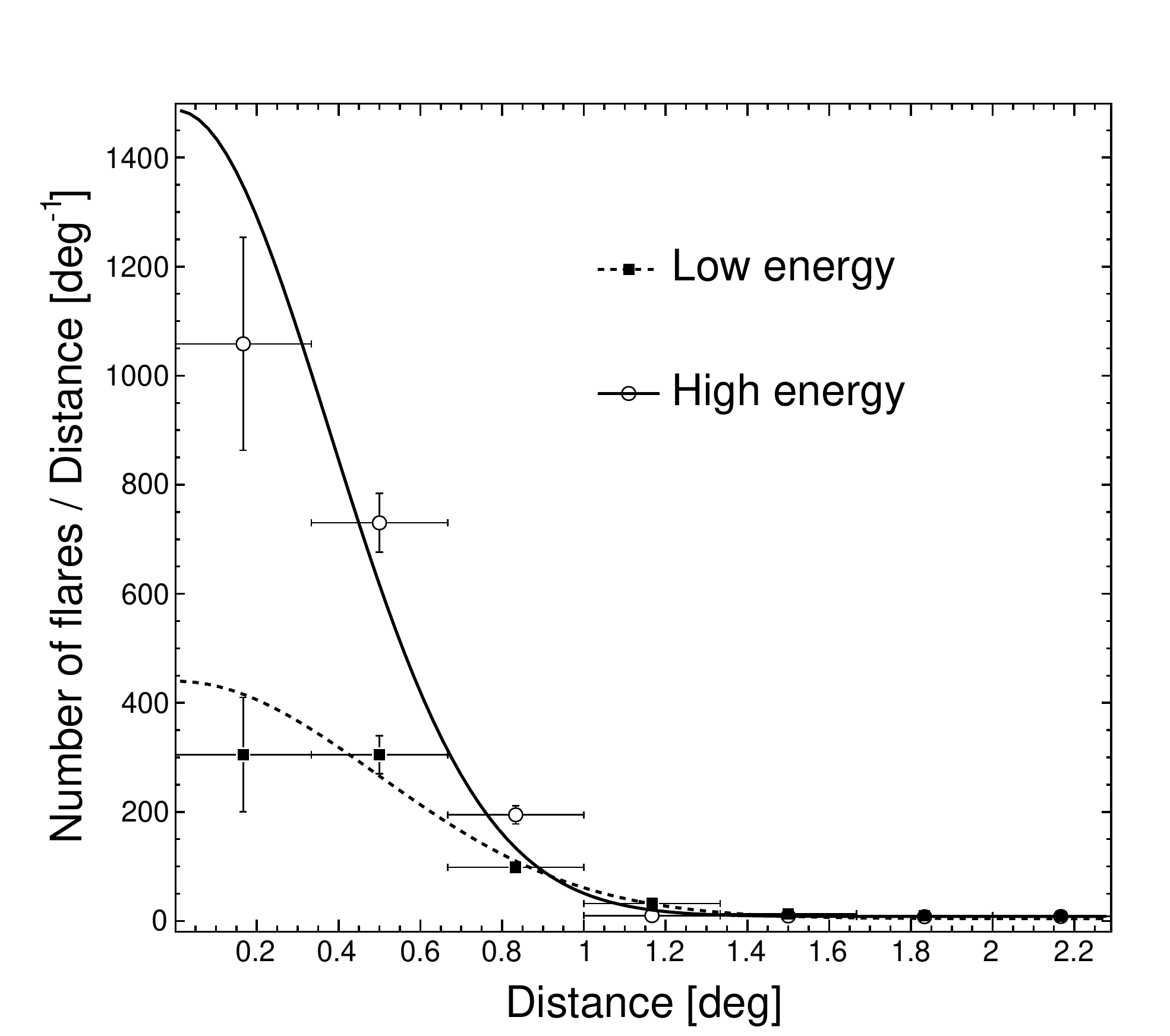} &
\end{tabular}
\end{center}
\caption{Distribution of angular distances between flares exceeding the significance criteria for source detection and known variable gamma-ray sources. The solid and dashed lines show the best-fit parametrizations for flares with a detection in the high energy band, and only in the low energy band, respectively (see text). }
\label{fig:angres}
\end{figure}

\begin{figure*}[ht]
\begin{center}
\begin{tabular}{cl}
 \includegraphics[width=1.0\textwidth]{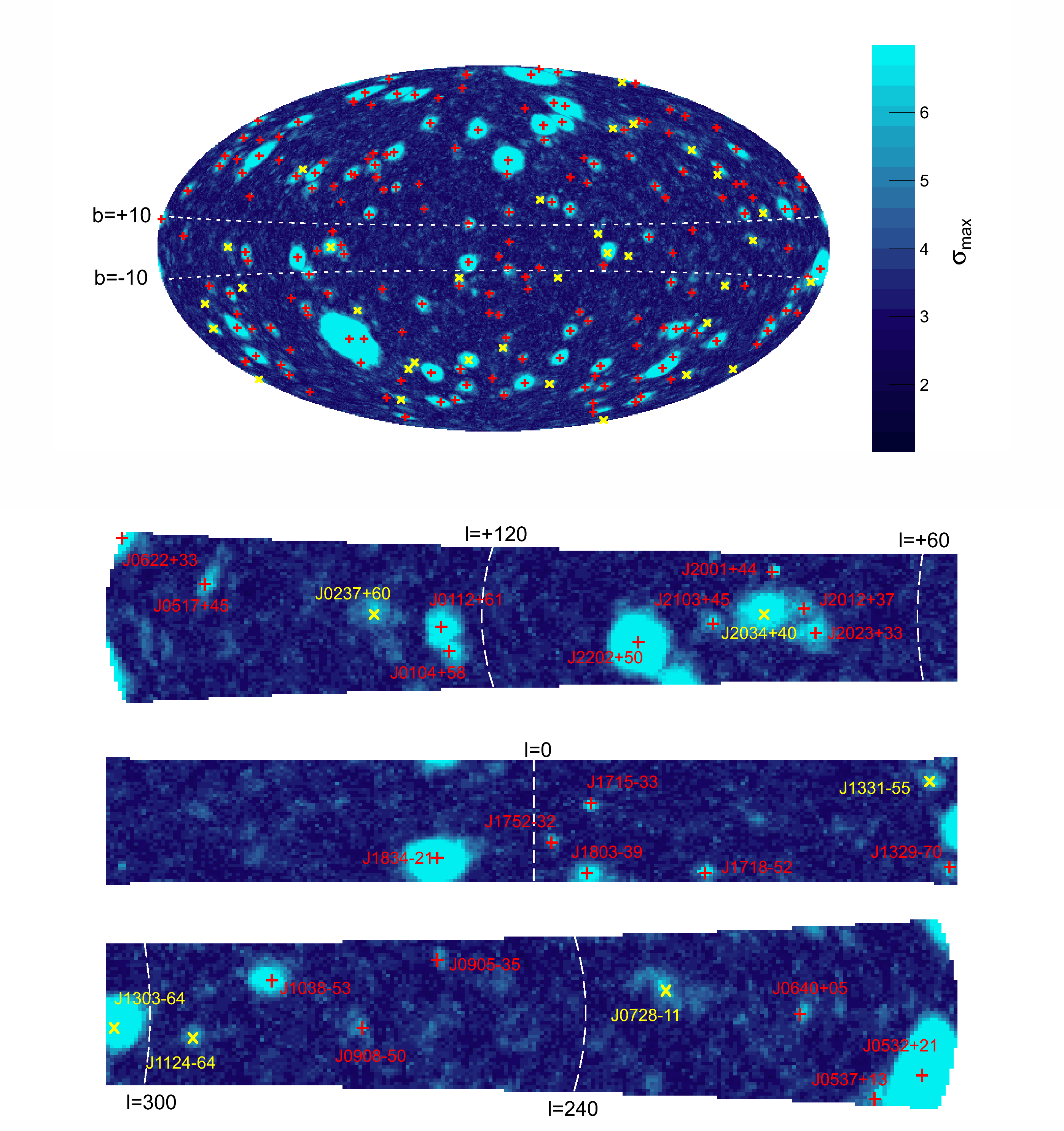} &
\end{tabular}
\end{center}
\caption{FAVA sources are shown in Galactic coordinates and a Hammer-Aitoff projection. Red crosses mark sources for which at least one flare was detected in the high-energy band. Sources that were detected only at low energies are marked by a yellow X. The colored image in the background shows the maximum significance $\sigma_{max}$ detected in each pixel either in the $>100$ MeV or $>800$ MeV energy bands during the first 47 months of \emph{Fermi} observations. The lower three panels show the region of Galactic latitude within 10$^\circ$ of the equator, the region enclosed by the dashed lines in the upper panel. }
\label{fig:cat}
\end{figure*}

To estimate the position accuracy achieved by the peak finding algorithm, we analyzed the distribution of flares around known flaring gamma-ray sources. As a source sample we chose the 249 sources flagged as most variable in the second \emph{Fermi}-LAT catalog  (a variability index $>$83.2 in \citealt{Nolan2012}). The resulting distribution of flares per solid angle is shown in Figure \ref{fig:angres}. We assumed that the reconstructed position of the peak finder follows a Gaussian distribution plus a constant background term from flares not associated with the sources in this representation. The best-fit model shown in Figure \ref{fig:angres} represents the data in good approximation. We proceeded to calculate the distances within which 68\% of the flares are contained in this parametrization. At low energies the radius is 0.8$^\circ$ and at high energies it is 0.6$^\circ$. We verified that no systematic offset is present in the position estimation in any coordinate direction.

Most variable gamma-ray sources, such as AGN or X-ray binaries, are known to have recurring flares. We therefore group the detected flares, and associate closely located flares to a single common flaring source. For this purpose we used a Minimum Spanning Tree (MST, \citealt{Nea2001}). We first group the flares detected at high energies because their positions are better determined. We build the MST for these flares and merge neighboring flares with a distance of less than 2$^\circ$ in the spanning tree. We proceed to associate flares detected only at low energies to the ones found at high energy if their distance is less than 3$^\circ$. Finally, we build the MST of the low-energy flares that were not associated and merge neighboring flares with a distance of less than 3$^\circ$ in the MST. 


\begin{figure*}[h!t]
\begin{center}
\begin{tabular}{cl}
 \includegraphics[width=0.9\textwidth]{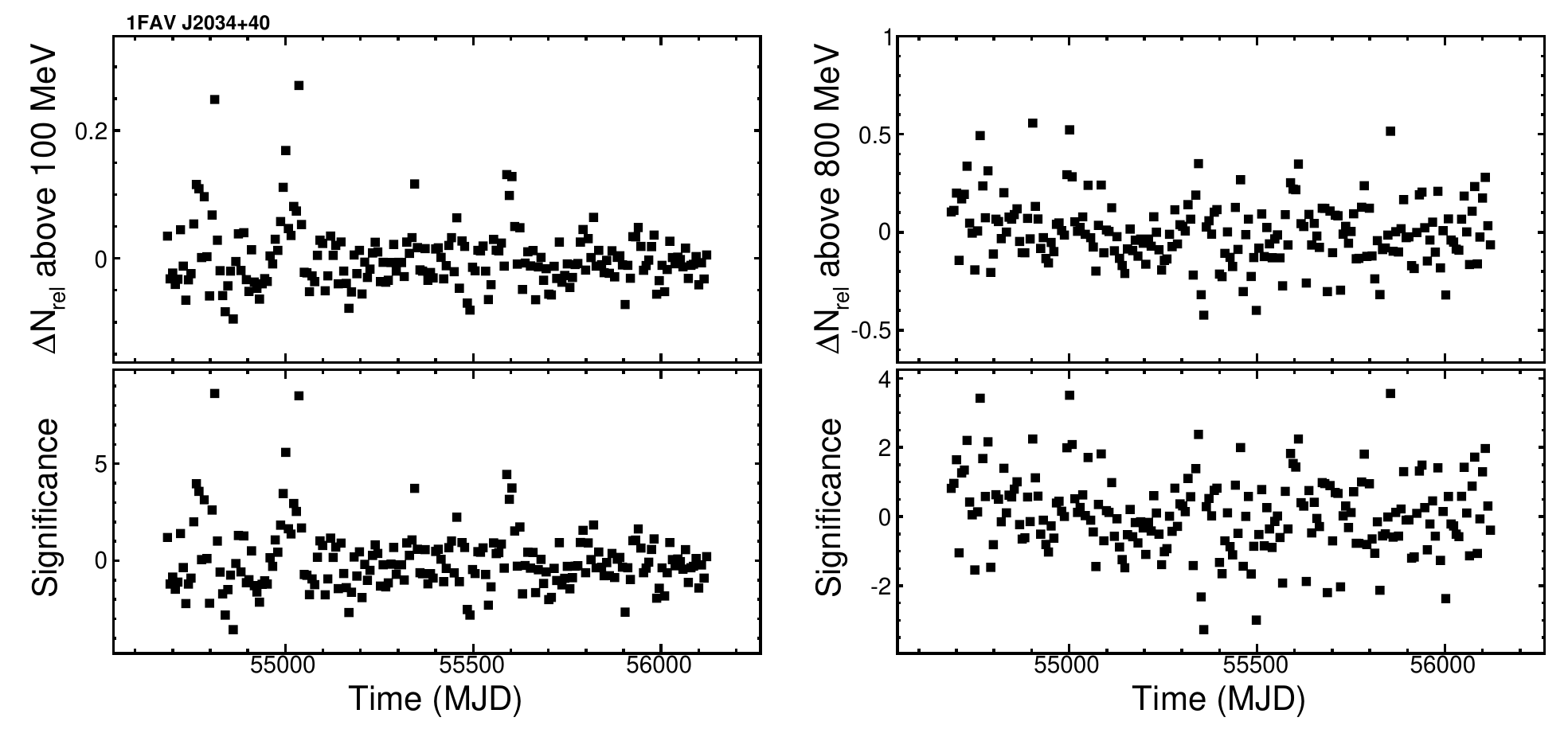} &
\end{tabular}
\end{center}
\caption{The top panels show the relative variations of counts $\Delta N_{rel} = (N - N^{exp}) / N^{exp}$ from the direction of Cyg X-3 (1FAV J2034+40) in weekly time bins, where $N$ is the number of measured counts and $N^{exp}$ is the number of expected counts from the average emission. Bottom panels show the significances of these variations. Left-hand plots are for photon energies above 100 MeV, the right-hand plots are for photon energies above 800 MeV.  }
\label{fig:lc}
\end{figure*}

The position of a FAVA source is found by averaging over the positions of its flares. If the source is detected at high energies we use only the positions of these flares due to their better positional accuracy; otherwise the low-energy flares are used. The position error of the source is obtained by propagating the positional errors of the flares included. In addition to this statistical error, there is a systematic error, that can arise from false associations of flares to a source, as well as the finite binning of the sky maps. We estimated this error to be smaller than $r_{sys}= 0.1^\circ$ by comparing the position of the FAVA sources to those in the list of variable 2FGL sources used previously. We assume $r_{sys}$ to be the systematic error on the source positions. 

A total of 215 sources are detected by FAVA. Out of these, 33 are detected at low energies only.  Flares related to negative flux variations from the average emission are found for 22 sources, often during periods of quiescent emission. All of the latter also showed positive flares. No source was found which flared only due to a negative flux variation.  Each FAVA source is referred to by its identification number composed of the right ascension in hours and minutes and the declination in degrees of the source (1FAV HHMM-DD). The positions of all sources in the sky are displayed in Figure \ref{fig:cat}. We produced light curves of relative flux variations with FAVA for all sources and made them publicly available online\footnote{https://www-glast.stanford.edu/pub\_data/585}. One example light curve is shown for the position of the high-mass X-ray binary Cyg X-3 in Figure \ref{fig:lc}.  

We looked for associations of FAVA sources with previously known variable LAT sources. We searched for counterparts within radius $R_s$, which is defined as the 99\% statistical error on the source position plus the systematic error $r_{sys}$. $R_s$ was deliberately chosen large,  to include all possible counterparts.  In cases for which more than one counterpart is found within $R_s$, we consider the closest one. The values of $R_s$ for each source and the found counterparts are listed in Table \ref{tab:cat}. We note that the associations were made purely on the basis of positional coincidence. We therefore caution that the associated sources should be considered as likely counterparts only. For a more confident source association temporal and spectral information need to be considered. Additionally, the positional localizations of the FAVA sources could be improved by analyzing each source individually with standard likelihood techniques. This is beyond the scope of this paper. The associations were assigned as follows:

\begin{itemize}
\item We searched for counterparts among the variable sources in the second  \emph{Fermi} source catalog (2FGL, \citealt{Nolan2012}). We restricted the search to the 458 2FGL sources that have a probability of less than 1\% of being constant on monthly time scales. We find a variable 2FGL source within the search radius for 170 of the FAVA sources. For those sources where no 2FGL counterpart was found we searched for association with sources in the first \emph{Fermi} source catalog (1FGL, \citealt{Abdo2010d}). The reason is that sources that flared only once at the beginning of the mission might be detected in the 1FGL but not in the 2FGL due to the increased integration time in the latter. We restricted the search to the 241 1FGL sources that have a probability of less than 1\% of being constant, finding an association for one source.
\item We searched for positional coincidences with \emph{Fermi}-LAT detected gamma-ray bursts\footnote{http://fermi.gsfc.nasa.gov/ssc/observations/types/grbs/} (GRBs) Even though GRBs have typical duration from a few seconds to minutes, their emission is sometimes bright enough to be detected over a time scale of 1 week. We find a GRB within the search radius for 4 FAVA sources. These FAVA sources flared only once and we have verified that the flare occurred during the week of the GRB outburst.
\item We searched for counterparts among LAT sources that were announced in Astronomer's Telegrams\footnote{https://www-glast.stanford.edu/cgi-bin/pub\_rapid ; http://www.asdc.asi.it/feratel (status November 2012)
} (ATels). These sources were found by the automated  sky processing  (ASP) used by the LAT Collaboration \citep{Atwood2009}. We found positional coincidences with 17 sources. 
\end{itemize}

We found LAT counterparts for 192 of the 215 FAVA sources. Most of the associated sources, 177, are AGN. All associations found at higher Galactic latitudes ($|b|>10$) belong to this class. Among the AGN associations 129 belong to the class of Flat Spectrum Radio Quasars (FSRQs) and 29 of them belong to the BL Lacertae (BL Lac) class. The number of gamma-ray emitting BL Lacs is approximately the same as the number of FSRQs, FSRQs therefore flare intrinsically more frequently in the LAT energy range. This is in agreement with the observation that FSRQs are more variable in gamma-rays on monthly time scales \citep{Ackermann2011b}. Three of the FAVA sources are associated to non-blazar AGN: two sources are associated to Narrow-Line Seyfert galaxies, which were recently found to be variable gamma-ray sources (1FAV J0320+41, \citealt{Donato2011}; 1FAV J0948+01, \citealt{Foschini2012}), and one source is found coincident with the radio galaxy NGC 1275 (1FAV J0320+41, \citealt{Kataoka2010}). The remaining associated sources are AGN of unknown type.





\section{Gamma-ray flares in the Galactic plane}

Of the 215 FAVA sources, 27 are detected at Galactic latitudes smaller than 10$^\circ$; their positions are shown in the lower panels of Figure \ref{fig:cat}. We found associations to previously known LAT sources for all of them. The low-latitude FAVA sources can be grouped into a few categories:
\begin{itemize}

\item Sources associated with Galactic sources: 4 sources coincide with variable 2FGL sources which are associated to Galactic sources. These are high-mass X-ray binaries Cyg X-3 (1FAV J308+41) and LSI +61 303 (1FAV J0237+60), the Crab Nebula (1FAV J0532+21), and the nova V407 Cyg (1FAV J2103+45). In addition, three sources are found coincident with LAT sources announced in ATels: the high-mass X-ray Binaries  PSRB 1259$-$63 (1FAV J1303$-$64, \citealt{Abdo2010e,Abdo2011b} ) and two sources that are likely associated to novas (1FAV J1752$-$32, \citealt{Cheung2012a}; 1FAV J0640+05, \citealt{Cheung2012b,CheungC.C.2012}).

\item Sources associated with blazars:  we find associations with variable 2FGL sources for 15 sources and to one 1FGL source  (1FAV J1124$-$64), which were  classified either as blazars or as AGN of uncertain type in the respective catalogs. Additionally, three sources were found coincident with sources announced in ATels and associated to blazars (1FAV J0537+13, \citealt{Orienti2012}; 1FAV J2202+50, \citealt{CipriniStefano2012}; 1FAV J1718$-$52, \citealt{Chomiuk2013}).  

\item Source with association of unknown type:  one source is associated to counterparts of unknown type announced in ATels (1FAV J1038$-$53, \citealt{Ciprini2012}). A compact radio sources is found coincident with its position.

\end{itemize}

Based on these associations 7 FAVA sources are located within the Milky Way. The source with the radio counterpart of unknown type might also be Galactic. Additionally, some of the made associations with blazars might turn out to be wrong. However, we can already infer statistically that most of the sources that we have not associated with Galactic sources are indeed extragalactic by calculating the number of expected extragalactic flares at low latitudes. The derivation relies on two assumptions:
\begin{enumerate}
\item The majority of the sources at high Galactic latitudes are extragalactic.
\item Extragalactic sources are isotropically distributed in the sky.
\end{enumerate}
We derive the number of extragalactic sources within 10$^\circ$ of the Galactic equator from the density of sources at latitudes greater than 30$^\circ$. After considering the difference in solid angle one expects 41 extragalactic sources at low latitudes. To take into account the reduced sensitivity for flare detection in the Galactic plane, we assigned random positions at low latitudes to the high-latitude sources. We determined the fraction of flares which would still be detected at the reduced sensitivity for each source. This results in an expectation of 24.6 variable extragalactic sources at Galactic latitudes smaller than 10$^{\circ}$.  The probability to detect 20 or fewer flares at low Galactic latitudes is 21\%. The 20 sources not associated with Galactic sources are therefore compatible with being all extragalactic. No more than 6 of them can have a Galactic origin at $>90$\% confidence.

We note that two gamma-ray binaries LS 5039  and 1FGL J1018.6-5856 \citep{Abdo2009b,Hadasch2012,FermiLATCollaboration2012} are not detected by FAVA . Their orbital periods of 3.9 d and 16.6 d, respectively, result in average weekly flux variations below the sensitivity for flare detection by FAVA. The X-ray binary Cyg X-1 is also not found by our analysis. The flare reported from this source by \citet{Sabatini2010} could not be confirmed by the LAT Collaboration\footnote{http://fermisky.blogspot.de/2010/03/lat-limit-on-cyg-x-1-during-reported.html}.

\section{Summary  \& Outlook}

We have presented the analysis tool FAVA, which searches for variable gamma-ray emitters in \emph{Fermi}-LAT data. We used FAVA to search for gamma-ray flares on weekly time scales over the entire sky. From these flares we derived a list of 215 flaring gamma-ray sources. A list of sources, their light curves and their associations are available online\footnote{https://www-glast.stanford.edu/pub\_data/585}. We searched for positional coincidences of these sources with previously known LAT sources, finding counterparts for 192 sources. We associated 177 sources with AGN and find that FSRQs flare more frequently than BL Lacs.

Twenty-seven of the FAVA sources are located at Galactic latitudes less than 10$^\circ$. We associated 7 of these to known Galactic sources. Among the remaining 20 sources, we found no evidence for new gamma-ray transients in our Galaxy. On the contrary, we showed that the majority of them  are probably extragalactic. For 19 sources we find positional coincidence with AGN. The one remaining source is associated to compact radio sources of unknown type. Future multi-waveband observations may reveal its nature.

No flare was detected from a pulsar other than the Crab Nebula. It remains puzzling why this source is the only one of its kind to exhibit long-term variability and flaring behavior. We cannot confirm the hypothesis reported by \citet{Neronov2012}, that gamma-ray variability might be common in young pulsars. We also detected no flares associated with previously undetected X-ray binary systems. These systems appear to be less efficient gamma-ray emitters than expected before the beginning of \emph{Fermi} observations  \citep{DubusG.2007}. We note that we detected no flares from the Galactic center region, which might have been expected if its gamma-ray emission was linked to accretion on the central black hole Sgr A* \citep{Aharonian2008}. 

In the future we plan to apply FAVA on different time scales, and scan the gamma-ray sky for short-term flares on time scales of a few hours, and for long-term flux variations of a few months. Furthermore, we intend to run the analysis  routinely to search for flares as soon as the LAT data are processed and sent to the {\it Fermi} Science Support Center. This will complement the ASP flare search currently used by the LAT Collaboration, and will help to alert the astrophysics community about gamma-ray flares in real time. 

\acknowledgments
Rolf B\"uhler acknowledges generous support from the \textit{Fermi} guest investigator program. The \textit{Fermi} LAT Collaboration acknowledges generous ongoing support
from a number of agencies and institutes that have supported both the
development and the operation of the LAT as well as scientific data analysis.
These include the National Aeronautics and Space Administration and the
Department of Energy in the United States, the Commissariat \`a l'Energie Atomique
and the Centre National de la Recherche Scientifique / Institut National de Physique
Nucl\'eaire et de Physique des Particules in France, the Agenzia Spaziale Italiana
and the Istituto Nazionale di Fisica Nucleare in Italy, the Ministry of Education,
Culture, Sports, Science and Technology (MEXT), High Energy Accelerator Research
Organization (KEK) and Japan Aerospace Exploration Agency (JAXA) in Japan, and
the K.~A.~Wallenberg Foundation, the Swedish Research Council and the
Swedish National Space Board in Sweden. Additional support for science analysis during the operations phase is gratefully
acknowledged from the Istituto Nazionale di Astrofisica in Italy and the Centre National d' Etudes Spatiales in France.

\bibliography{fava}{}

\begin{thebibliography}{48}
\expandafter\ifx\csname natexlab\endcsname\relax\def\natexlab#1{#1}\fi

\bibitem[{Abdo {et~al.}(2010{\natexlab{a}})Abdo, Parent, Grove, Caliandro,
  Roberts, Johnston, \& Chernyakova}]{Abdo2010e}
Abdo, A.~A., Parent, D., Grove, J.~E., {et~al.} 2010{\natexlab{a}}, The
  Astronomer's Telegram, 3085

\bibitem[{Abdo {et~al.}(2009{\natexlab{a}})Abdo, Ackermann, Ajello, Atwood,
  Axelsson, Baldini, Ballet, Barbiellini, Bastieri, Baughman, Bechtol,
  Bellazzini, Berenji, Blandford, Bloom, Bonamente, Borgland, Bregeon, Brez,
  Brigida, Bruel, Burnett, Buson, Caliandro, Cameron, Caraveo, Casandjian,
  Cavazzuti, Cecchi, \c{C}elik, Chaty, Chekhtman, Cheung, Chiang, Ciprini,
  Claus, Cohen-Tanugi, Cominsky, Conrad, Corbel, Corbet, Cutini, Dermer,
  de~Angelis, de~Palma, Digel, {do Couto e Silva}, Drell, Dubois, Dubus,
  Dumora, Farnier, Favuzzi, Fegan, Focke, Fortin, Frailis, Fukazawa, Funk,
  Fusco, Gargano, Gasparrini, Gehrels, Germani, Giebels, Giglietto, Giordano,
  Glanzman, Godfrey, Grenier, Grondin, Grove, Guillemot, Guiriec, Hanabata,
  Harding, Hayashida, Hays, Hill, Horan, Hughes, Jackson, J\'{o}hannesson,
  Johnson, Johnson, Johnson, Kamae, Katagiri, Kataoka, Kawai, Kerr,
  Kn\"{o}dlseder, Kocian, Kuehn, Kuss, Lande, Larsson, Latronico,
  Lemoine-Goumard, Longo, Loparco, Lott, Lovellette, Lubrano, Madejski, Makeev,
  Marelli, Mazziotta, McEnery, Meurer, Michelson, Mitthumsiri, Mizuno, Moiseev,
  Monte, Monzani, Morselli, Moskalenko, Murgia, Nolan, Norris, Nuss, Ohsugi,
  Omodei, Orlando, Ormes, Ozaki, Paneque, Panetta, Parent, Pelassa, Pepe,
  Pesce-Rollins, Piron, Porter, Rain\`{o}, Rando, Ray, Razzano, Rea, Reimer,
  Reimer, Reposeur, Ritz, Rochester, Rodriguez, Romani, Roth, Ryde,
  Sadrozinski, Sanchez, Sander, {Saz Parkinson}, Scargle, Sgr\`{o},
  Sierpowska-Bartosik, Siskind, Smith, Smith, Spandre, Spinelli, Strickman,
  Suson, Tajima, Takahashi, Takahashi, Tanaka, Tanaka, Thayer, Thompson,
  Tibaldo, Torres, Tosti, Tramacere, Uchiyama, Usher, Vasileiou, Venter,
  Vilchez, Vitale, Waite, Wallace, Wang, Winer, Wood, Ylinen, \&
  Ziegler}]{Abdo2009b}
Abdo, A.~A., Ackermann, M., Ajello, M., {et~al.} 2009{\natexlab{a}}, The
  Astrophysical Journal, 706, L56

\bibitem[{Abdo {et~al.}(2009{\natexlab{b}})Abdo, Ackermann, Ajello, Axelsson,
  Baldini, Ballet, Barbiellini, Bastieri, Baughman, Bechtol, Bellazzini,
  Berenji, Blandford, Bloom, Bonamente, Borgland, Brez, Brigida, Bruel,
  Burnett, Buson, Caliandro, Cameron, Caraveo, Casandjian, Cecchi, Celik,
  Chaty, Cheung, Chiang, Ciprini, Claus, Cohen-Tanugi, Cominsky, Conrad,
  Corbel, Corbet, Dermer, de~Palma, Digel, {do Couto e Silva}, Drell, Dubois,
  Dubus, Dumora, Farnier, Favuzzi, Fegan, Focke, Fortin, Frailis, Fusco,
  Gargano, Gehrels, Germani, Giavitto, Giebels, Giglietto, Giordano, Glanzman,
  Godfrey, Grenier, Grondin, Grove, Guillemot, Guiriec, Hanabata, Harding,
  Hayashida, Hays, Hill, Hjalmarsdotter, Horan, Hughes, Jackson,
  J\'{o}hannesson, Johnson, Johnson, Johnson, Kamae, Katagiri, Kawai, Kerr,
  Kn\"{o}dlseder, Kocian, Koerding, Kuss, Lande, Latronico, Lemoine-Goumard,
  Longo, Loparco, Lott, Lovellette, Lubrano, Madejski, Makeev, Marchand,
  Marelli, Max-Moerbeck, Mazziotta, McColl, McEnery, Meurer, Michelson,
  Migliari, Mitthumsiri, Mizuno, Monte, Monzani, Morselli, Moskalenko, Murgia,
  Nolan, Norris, Nuss, Ohsugi, Omodei, Ong, Ormes, Paneque, Parent, Pelassa,
  Pepe, Pesce-Rollins, Piron, Pooley, Porter, Pottschmidt, Rain\`{o}, Rando,
  Ray, Razzano, Rea, Readhead, Reimer, Reimer, Richards, Rochester, Rodriguez,
  Rodriguez, Romani, Ryde, Sadrozinski, Sander, {Saz Parkinson}, Sgr\`{o},
  Siskind, Smith, Smith, Spinelli, Starck, Stevenson, Strickman, Suson,
  Takahashi, Tanaka, Thayer, Thompson, Tibaldo, Tomsick, Torres, Tosti,
  Tramacere, Uchiyama, Usher, Vasileiou, Vilchez, Vitale, Waite, Wang, Wilms,
  Winer, Wood, Ylinen, \& Ziegler}]{Abdo2009}
---. 2009{\natexlab{b}}, Science (New York, N.Y.), 326, 1512

\bibitem[{Abdo {et~al.}(2010{\natexlab{b}})Abdo, Ackermann, Ajello, Allafort,
  Antolini, Atwood, Axelsson, Baldini, Ballet, Barbiellini, Bastieri, Baughman,
  Bechtol, Bellazzini, Belli, Berenji, Bisello, Blandford, Bloom, Bonamente,
  Bonnell, Borgland, Bouvier, Bregeon, Brez, Brigida, Bruel, Burnett, Busetto,
  Buson, Caliandro, Cameron, Campana, Canadas, Caraveo, Carrigan, Casandjian,
  Cavazzuti, Ceccanti, Cecchi, \c{C}elik, Charles, Chekhtman, Cheung, Chiang,
  Cillis, Ciprini, Claus, Cohen-Tanugi, Conrad, Corbet, Davis, DeKlotz, den
  Hartog, Dermer, de~Angelis, de~Luca, de~Palma, Digel, Dormody, {do Couto e
  Silva}, Drell, Dubois, Dumora, Fabiani, Farnier, Favuzzi, Fegan, Ferrara,
  Focke, Fortin, Frailis, Fukazawa, Funk, Fusco, Gargano, Gasparrini, Gehrels,
  Germani, Giavitto, Giebels, Giglietto, Giommi, Giordano, Giroletti, Glanzman,
  Godfrey, Grenier, Grondin, Grove, Guillemot, Guiriec, Gustafsson, Hadasch,
  Hanabata, Harding, Hayashida, Hays, Healey, Hill, Horan, Hughes, Iafrate,
  J\'{o}hannesson, Johnson, Johnson, Johnson, Johnson, Kamae, Katagiri,
  Kataoka, Kawai, Kerr, Kn\"{o}dlseder, Kocevski, Kuss, Lande, Landriu,
  Latronico, Lee, Lemoine-Goumard, Lionetto, {Llena Garde}, Longo, Loparco,
  Lott, Lovellette, Lubrano, Madejski, Makeev, Marangelli, Marelli, Massaro,
  Mazziotta, McConville, McEnery, Michelson, Minuti, Mitthumsiri, Mizuno,
  Moiseev, Mongelli, Monte, Monzani, Moretti, Morselli, Moskalenko, Murgia,
  Nakajima, Nakamori, Naumann-Godo, Nolan, Norris, Nuss, Ohno, Ohsugi, Omodei,
  Orlando, Ormes, Ozaki, Paccagnella, Paneque, Panetta, Parent, Pelassa, Pepe,
  Pesce-Rollins, Pinchera, Piron, Porter, Poupard, Rain\`{o}, Rando, Ray,
  Razzano, Razzaque, Rea, Reimer, Reimer, Reposeur, Ripken, Ritz, Rochester,
  Rodriguez, Romani, Roth, Sadrozinski, Salvetti, Sanchez, Sander, {Saz
  Parkinson}, Scargle, Schalk, Scolieri, Sgr\`{o}, Shaw, Siskind, Smith, Smith,
  Spandre, Spinelli, Starck, Stephens, Striani, Strickman, Strong, Suson,
  Tajima, Takahashi, Takahashi, Tanaka, Thayer, Thayer, Thompson, Tibaldo,
  Tibolla, Tinebra, Torres, Tosti, Tramacere, Uchiyama, Usher, {Van Etten},
  Vasileiou, Vilchez, Vitale, Waite, Wallace, Wang, Watters, Winer, Wood, Yang,
  Ylinen, \& Ziegler}]{Abdo2010d}
---. 2010{\natexlab{b}}, The Astrophysical Journal Supplement Series, 188, 405

\bibitem[{Abdo {et~al.}(2010{\natexlab{c}})Abdo, Ackermann, Ajello, Allafort,
  Baldini, Ballet, Barbiellini, Bastieri, Bechtol, Bellazzini, Berenji,
  Blandford, Bonamente, Borgland, Bouvier, Brandt, Bregeon, Brez, Brigida,
  Bruel, Buehler, Burnett, Caliandro, Cameron, Caraveo, Carrigan, Casandjian,
  Cecchi, \c{C}elik, Chaty, Chekhtman, Cheung, Chiang, Ciprini, Claus,
  Cohen-Tanugi, Cominsky, Conrad, Dermer, de~Palma, Digel, Silva, Drell,
  Dubois, Dumora, Favuzzi, Fegan, Ferrara, Frailis, Fukazawa, Fusco, Gargano,
  Gehrels, Germani, Giglietto, Giordano, Godfrey, Grenier, Grondin, Grove,
  Guillemot, Guiriec, Hadasch, Hanabata, Harding, Hayashida, Hays, Hill, Horan,
  Hughes, Itoh, Jackson, J\'{o}hannesson, Johnson, Johnson, Kamae, Katagiri,
  Kataoka, Kerr, Kn\"{o}dlseder, Kuss, Lande, Latronico, Lee, Lemoine-Goumard,
  Livingstone, {Llena Garde}, Longo, Loparco, Lovellette, Lubrano, Makeev,
  Mazziotta, McEnery, Mehault, Michelson, Mitthumsiri, Mizuno, Moiseev, Monte,
  Monzani, Morselli, Moskalenko, Murgia, Nakamori, Naumann-Godo, Nolan, Norris,
  Nuss, Ohsugi, Okumura, Omodei, Orlando, Ormes, Ozaki, Panetta, Parent,
  Pelassa, Pepe, Pesce-Rollins, Piron, Porter, Rain\`{o}, Rando, Razzano,
  Reimer, Reimer, Reposeur, Rodriguez, Romani, Roth, Sadrozinski, Sander,
  Parkinson, Scargle, Sgr\`{o}, Siskind, Smith, Smith, Spandre, Spinelli,
  Strickman, Suson, Takahashi, Takahashi, Tanaka, Thayer, Thayer, Thompson,
  Tibaldo, Tibolla, Torres, Tosti, Tramacere, Uchiyama, Usher, Vandenbroucke,
  Vasileiou, Vilchez, Vitale, Waite, Wallace, Wang, Winer, Wood, Yang, Ylinen,
  \& Ziegler}]{Abdo2010a}
---. 2010{\natexlab{c}}, The Astrophysical Journal, 723, 649

\bibitem[{Abdo {et~al.}(2010{\natexlab{d}})Abdo, Ackermann, Ajello, Atwood,
  Baldini, Ballet, Barbiellini, Bastieri, Bechtol, Bellazzini, Berenji,
  Blandford, Bloom, Bonamente, Borgland, Bouvier, Brandt, Bregeon, Brez,
  Brigida, Bruel, Buehler, Burnett, Buson, Caliandro, Cameron, Caraveo,
  Carrigan, Casandjian, Cecchi, Celik, Charles, Chaty, Chekhtman, Cheung,
  Chiang, Ciprini, Claus, Cohen-Tanugi, Conrad, Corbel, Corbet, DeCesar, den
  Hartog, Dermer, de~Palma, Digel, Donato, {do Couto e Silva}, Drell, Dubois,
  Dubus, Dumora, Favuzzi, Fegan, Ferrara, Fortin, Frailis, Fuhrmann, Fukazawa,
  Funk, Fusco, Gargano, Gasparrini, Gehrels, Germani, Giglietto, Giordano,
  Giroletti, Glanzman, Godfrey, Grenier, Grondin, Grove, Guiriec, Hadasch,
  Harding, Hayashida, Hays, Healey, Hill, Horan, Hughes, Itoh, Jean,
  J\'{o}hannesson, Johnson, Johnson, Johnson, Johnson, Kamae, Katagiri,
  Kataoka, Kerr, Kn\"{o}dlseder, Koerding, Kuss, Lande, Latronico, Lee,
  Lemoine-Goumard, Garde, Longo, Loparco, Lott, Lovellette, Lubrano, Makeev,
  Mazziotta, McConville, McEnery, Mehault, Michelson, Mizuno, Moiseev, Monte,
  Monzani, Morselli, Moskalenko, Murgia, Nakamori, Naumann-Godo, Nestoras,
  Nolan, Norris, Nuss, Ohno, Ohsugi, Okumura, Omodei, Orlando, Ormes, Ozaki,
  Paneque, Panetta, Parent, Pelassa, Pepe, Pesce-Rollins, Piron, Porter,
  Rain\`{o}, Rando, Ray, Razzano, Razzaque, Rea, Reimer, Reimer, Reposeur,
  Ripken, Ritz, Romani, Roth, Sadrozinski, Sander, Parkinson, Scargle,
  Schinzel, Sgr\`{o}, Shaw, Siskind, Smith, Smith, Sokolovsky, Spandre,
  Spinelli, Stawarz, Strickman, Suson, Takahashi, Takahashi, Tanaka, Tanaka,
  Thayer, Thayer, Thompson, Tibaldo, Torres, Tosti, Tramacere, Uchiyama, Usher,
  Vandenbroucke, Vasileiou, Vilchez, Vitale, Waite, Wallace, Wang, Winer,
  Wolff, Wood, Yang, Ylinen, Ziegler, Maehara, Nishiyama, Kabashima, Bach,
  Bower, Falcone, Forster, Henden, Kawabata, Koubsky, Mukai, Nelson, Oates,
  Sakimoto, Sasada, Shenavrin, Shore, Skinner, Sokoloski, Stroh, Tatarnikov,
  Uemura, Wahlgren, \& Yamanaka}]{Abdo2010}
---. 2010{\natexlab{d}}, Science (New York, N.Y.), 329, 817

\bibitem[{Abdo {et~al.}(2010{\natexlab{e}})Abdo, Ackermann, Ajello, Atwood,
  Baldini, Ballet, Barbiellini, Bastieri, Baughman, Bechtol, Bellazzini,
  Berenji, Blandford, Bloom, Bonamente, Borgland, Bregeon, Brez, Brigida,
  Bruel, Burnett, Buson, Caliandro, Cameron, Caraveo, Casandjian, Cavazzuti,
  Cecchi, \c{C}elik, Charles, Chekhtman, Cheung, Chiang, Ciprini, Claus,
  Cohen-Tanugi, Cominsky, Conrad, Cutini, Dermer, de~Angelis, de~Palma, Digel,
  {Di Bernardo}, e~Silva, Drell, Drlica-Wagner, Dubois, Dumora, Farnier,
  Favuzzi, Fegan, Focke, Fortin, Frailis, Fukazawa, Funk, Fusco, Gaggero,
  Gargano, Gasparrini, Gehrels, Germani, Giebels, Giglietto, Giommi, Giordano,
  Glanzman, Godfrey, Grenier, Grondin, Grove, Guillemot, Guiriec, Gustafsson,
  Hanabata, Harding, Hayashida, Hughes, Itoh, Jackson, J\'{o}hannesson,
  Johnson, Johnson, Johnson, Johnson, Kamae, Katagiri, Kataoka, Kawai, Kerr,
  Kn\"{o}dlseder, Kocian, Kuehn, Kuss, Lande, Latronico, Lemoine-Goumard,
  Longo, Loparco, Lott, Lovellette, Lubrano, Madejski, Makeev, Mazziotta,
  McConville, McEnery, Meurer, Michelson, Mitthumsiri, Mizuno, Moiseev, Monte,
  Monzani, Morselli, Moskalenko, Murgia, Nolan, Norris, Nuss, Ohsugi, Omodei,
  Orlando, Ormes, Paneque, Panetta, Parent, Pelassa, Pepe, Pesce-Rollins,
  Piron, Porter, Rain\`{o}, Rando, Razzano, Reimer, Reimer, Reposeur, Ritz,
  Rochester, Rodriguez, Roth, Ryde, Sadrozinski, Sanchez, Sander, Parkinson,
  Scargle, Sellerholm, Sgr\`{o}, Shaw, Siskind, Smith, Smith, Spandre,
  Spinelli, Starck, Strickman, Strong, Suson, Tajima, Takahashi, Takahashi,
  Tanaka, Thayer, Thayer, Thompson, Tibaldo, Torres, Tosti, Tramacere,
  Uchiyama, Usher, Vasileiou, Vilchez, Vitale, Waite, Wang, Winer, Wood,
  Ylinen, \& Ziegler}]{Abdo2010c}
---. 2010{\natexlab{e}}, Physical Review Letters, 104

\bibitem[{Abdo {et~al.}(2010{\natexlab{f}})Abdo, Ackermann, Ajello, Antolini,
  Baldini, Ballet, Barbiellini, Bastieri, Baughman, Bechtol, Bellazzini,
  Berenji, Blandford, Bloom, Bonamente, Borgland, Bouvier, Bregeon, Brez,
  Brigida, Bruel, Burnett, Buson, Caliandro, Cameron, Caraveo, Carrigan,
  Casandjian, Cavazzuti, Cecchi, \c{C}elik, Charles, Chekhtman, Cheung, Chiang,
  Ciprini, Claus, Cohen-Tanugi, Conrad, Costamante, Cutini, Dermer, de~Angelis,
  de~Palma, {do Couto e Silva}, Drell, Dubois, Dumora, Farnier, Favuzzi, Fegan,
  Focke, Fukazawa, Funk, Fusco, Gargano, Gasparrini, Gehrels, Germani,
  Giglietto, Giommi, Giordano, Glanzman, Godfrey, Grenier, Grove, Guiriec,
  Hadasch, Hayashida, Hays, Healey, Horan, Hughes, Itoh, J\'{o}hannesson,
  Johnson, Johnson, Johnson, Kamae, Katagiri, Kataoka, Kawai, Kn\"{o}dlseder,
  Kuss, Lande, Latronico, Lee, Lemoine-Goumard, Garde, Longo, Loparco, Lott,
  Lovellette, Lubrano, Madejski, Makeev, Mazziotta, McConville, McEnery,
  Meurer, Michelson, Mitthumsiri, Mizuno, Monte, Monzani, Morselli, Moskalenko,
  Murgia, Nolan, Norris, Nuss, Ohsugi, Omodei, Orlando, Ormes, Ozaki, Paneque,
  Panetta, Parent, Pelassa, Pepe, Pesce-Rollins, Piron, Porter, Rain\`{o},
  Rando, Razzano, Reimer, Reimer, Ritz, Rochester, Rodriguez, Romani, Roth,
  Sadrozinski, Sander, Parkinson, Scargle, Sgr\`{o}, Shaw, Smith, Spandre,
  Spinelli, Starck, Strickman, Strong, Suson, Tajima, Takahashi, Takahashi,
  Tanaka, Thayer, Thayer, Thompson, Tibaldo, Torres, Tosti, Tramacere,
  Uchiyama, Usher, Vasileiou, Vilchez, Vitale, Waite, Wang, Winer, Wood, Yang,
  Ylinen, \& Ziegler}]{Abdo2010b}
---. 2010{\natexlab{f}}, The Astrophysical Journal, 720, 435

\bibitem[{Abdo {et~al.}(2011{\natexlab{a}})Abdo, Ackermann, Ajello, Allafort,
  Ballet, Barbiellini, Bastieri, Bechtol, Bellazzini, Berenji, Blandford,
  Bonamente, Borgland, Bregeon, Brigida, Bruel, Buehler, Buson, Caliandro,
  Cameron, Camilo, Caraveo, Cecchi, Charles, Chaty, Chekhtman, Chernyakova,
  Cheung, Chiang, Ciprini, Claus, Cohen-Tanugi, Cominsky, Corbel, Cutini,
  D'Ammando, de~Angelis, den Hartog, de~Palma, Dermer, Digel, {do Couto e
  Silva}, Dormody, Drell, Drlica-Wagner, Dubois, Dubus, Dumora, Enoto,
  Espinoza, Favuzzi, Fegan, Ferrara, Focke, Fortin, Fukazawa, Funk, Fusco,
  Gargano, Gasparrini, Gehrels, Germani, Giglietto, Giommi, Giordano,
  Giroletti, Glanzman, Godfrey, Grenier, Grondin, Grove, Grundstrom, Guiriec,
  Gwon, Hadasch, Harding, Hayashida, Hays, J\'{o}hannesson, Johnson, Johnson,
  Johnston, Kamae, Katagiri, Kataoka, Keith, Kerr, Kn\"{o}dlseder, Kramer,
  Kuss, Lande, Lee, Lemoine-Goumard, Longo, Loparco, Lovellette, Lubrano,
  Manchester, Marelli, Mazziotta, Michelson, Mitthumsiri, Mizuno, Moiseev,
  Monte, Monzani, Morselli, Moskalenko, Murgia, Nakamori, Naumann-Godo,
  Neronov, Nolan, Norris, Noutsos, Nuss, Ohsugi, Okumura, Omodei, Orlando,
  Paneque, Parent, Pesce-Rollins, Pierbattista, Piron, Porter, Possenti,
  Rain\`{o}, Rando, Ray, Razzano, Razzaque, Reimer, Reimer, Reposeur, Ritz,
  Sadrozinski, Scargle, Sgr\`{o}, Shannon, Siskind, Smith, Spandre, Spinelli,
  Strickman, Suson, Takahashi, Tanaka, Thayer, Thayer, Thompson, Thorsett,
  Tibaldo, Tibolla, Torres, Tosti, Troja, Uchiyama, Usher, Vandenbroucke,
  Vasileiou, Vianello, Vitale, Waite, Wang, Winer, Wolff, Wood, Wood, Yang,
  Ziegler, \& Zimmer}]{Abdo2011b}
---. 2011{\natexlab{a}}, The Astrophysical Journal, 736, L11

\bibitem[{Abdo {et~al.}(2011{\natexlab{b}})Abdo, Ackermann, Ajello, Baldini,
  Ballet, Barbiellini, Bastieri, Bechtol, Bellazzini, Berenji, Bonamente,
  Borgland, Bouvier, Bregeon, Brez, Brigida, Bruel, Buehler, Buson, Caliandro,
  Cameron, Caraveo, Casandjian, Cecchi, Charles, Chekhtman, Chiang, Ciprini,
  Claus, Cohen-Tanugi, Conrad, Cutini, de~Angelis, de~Palma, Dermer, Digel, {do
  Couto e Silva}, Drell, Dubois, Favuzzi, Fegan, Focke, Fortin, Frailis, Funk,
  Fusco, Gargano, Gasparrini, Gehrels, Germani, Giglietto, Giordano, Giroletti,
  Glanzman, Godfrey, Grenier, Grillo, Guiriec, Hadasch, Hays, Hughes, Iafrate,
  J\'{o}hannesson, Johnson, Johnson, Kamae, Katagiri, Kataoka, Kn\"{o}dlseder,
  Kuss, Lande, Latronico, Lee, Lionetto, Longo, Loparco, Lott, Lovellette,
  Lubrano, Makeev, Mazziotta, McEnery, Mehault, Michelson, Mitthumsiri, Mizuno,
  Moiseev, Monte, Monzani, Morselli, Moskalenko, Murgia, Nakamori,
  Naumann-Godo, Nolan, Norris, Nuss, Ohsugi, Okumura, Omodei, Orlando, Ormes,
  Ozaki, Paneque, Pelassa, Pesce-Rollins, Pierbattista, Piron, Porter,
  Rain\`{o}, Rando, Razzano, Reimer, Reimer, Reposeur, Ritz, Sadrozinski,
  Schalk, Sgr\`{o}, Share, Siskind, Smith, Spandre, Spinelli, Strickman,
  Strong, Takahashi, Tanaka, Thayer, Thayer, Thompson, Tibaldo, Torres, Tosti,
  Tramacere, Troja, Uchiyama, Usher, Vandenbroucke, Vasileiou, Vianello,
  Vilchez, Vitale, Vladimirov, Waite, Wang, Winer, Wood, Yang, \&
  Ziegler}]{Abdo2011a}
---. 2011{\natexlab{b}}, The Astrophysical Journal, 734, 116

\bibitem[{Abdo {et~al.}(2011{\natexlab{c}})Abdo, Ackermann, Ajello, Allafort,
  Baldini, Ballet, Barbiellini, Bastieri, Bechtol, Bellazzini, Berenji,
  Blandford, Bloom, Bonamente, Borgland, Bouvier, Brandt, Bregeon, Brez,
  Brigida, Bruel, Buehler, Buson, Caliandro, Cameron, Cannon, Caraveo,
  Casandjian, \c{C}elik, Charles, Chekhtman, Cheung, Chiang, Ciprini, Claus,
  Cohen-Tanugi, Costamante, Cutini, D'Ammando, Dermer, de~Angelis, de~Luca,
  de~Palma, Digel, {do Couto e Silva}, Drell, Drlica-Wagner, Dubois, Dumora,
  Favuzzi, Fegan, Ferrara, Focke, Fortin, Frailis, Fukazawa, Funk, Fusco,
  Gargano, Gasparrini, Gehrels, Germani, Giglietto, Giordano, Giroletti,
  Glanzman, Godfrey, Grenier, Grondin, Grove, Guiriec, Hadasch, Hanabata,
  Harding, Hayashi, Hayashida, Hays, Horan, Itoh, J\'{o}hannesson, Johnson,
  Johnson, Khangulyan, Kamae, Katagiri, Kataoka, Kerr, Kn\"{o}dlseder, Kuss,
  Lande, Latronico, Lee, Lemoine-Goumard, Longo, Loparco, Lubrano, Madejski,
  Makeev, Marelli, Mazziotta, McEnery, Michelson, Mitthumsiri, Mizuno, Moiseev,
  Monte, Monzani, Morselli, Moskalenko, Murgia, Nakamori, Naumann-Godo, Nolan,
  Norris, Nuss, Ohsugi, Okumura, Omodei, Ormes, Ozaki, Paneque, Parent,
  Pelassa, Pepe, Pesce-Rollins, Pierbattista, Piron, Porter, Rain\`{o}, Rando,
  Ray, Razzano, Reimer, Reimer, Reposeur, Ritz, Romani, Sadrozinski, Sanchez,
  {Saz Parkinson}, Scargle, Schalk, Sgr\`{o}, Siskind, Smith, Spandre,
  Spinelli, Strickman, Suson, Takahashi, Takahashi, Tanaka, Thayer, Thompson,
  Tibaldo, Torres, Tosti, Tramacere, Troja, Uchiyama, Vandenbroucke, Vasileiou,
  Vianello, Vitale, Wang, Wood, Yang, \& Ziegler}]{Abdo2011}
---. 2011{\natexlab{c}}, Science (New York, N.Y.), 331, 739

\bibitem[{Ackermann {et~al.}(2011{\natexlab{a}})Ackermann, Ajello, Allafort,
  Angelakis, Axelsson, Baldini, Ballet, Barbiellini, Bastieri, Bellazzini,
  Berenji, Blandford, Bloom, Bonamente, Borgland, Bouvier, Bregeon, Brez,
  Brigida, Bruel, Buehler, Buson, Caliandro, Cameron, Cannon, Caraveo,
  Casandjian, Cavazzuti, Cecchi, Charles, Chekhtman, Cheung, Ciprini, Claus,
  Cohen-Tanugi, Cutini, de~Palma, Dermer, {do Couto e Silva}, Drell, Dubois,
  Dumora, Escande, Favuzzi, Fegan, Focke, Fortin, Frailis, Fuhrmann, Fukazawa,
  Fusco, Gargano, Gasparrini, Gehrels, Giglietto, Giommi, Giordano, Giroletti,
  Glanzman, Godfrey, Grandi, Grenier, Guiriec, Hadasch, Hayashida, Hays,
  Healey, J\'{o}hannesson, Johnson, Kamae, Katagiri, Kataoka, Kn\"{o}dlseder,
  Kuss, Lande, Lee, Longo, Loparco, Lott, Lovellette, Lubrano, Makeev,
  Max-Moerbeck, Mazziotta, McEnery, Mehault, Michelson, Mizuno, Monte, Monzani,
  Morselli, Moskalenko, Murgia, Naumann-Godo, Nishino, Nolan, Norris, Nuss,
  Ohsugi, Okumura, Omodei, Orlando, Ormes, Ozaki, Paneque, Pavlidou, Pelassa,
  Pepe, Pesce-Rollins, Pierbattista, Piron, Porter, Rain\`{o}, Razzano,
  Readhead, Reimer, Reimer, Richards, Romani, Sadrozinski, Scargle, Sgr\`{o},
  Siskind, Smith, Spandre, Spinelli, Strickman, Suson, Takahashi, Tanaka,
  Taylor, Thayer, Thayer, Thompson, Torres, Tosti, Tramacere, Troja,
  Vandenbroucke, Vianello, Vitale, Waite, Wang, Winer, Wood, Yang, \&
  Ziegler}]{Ackermann2011a}
Ackermann, M., Ajello, M., Allafort, A., {et~al.} 2011{\natexlab{a}}, The
  Astrophysical Journal, 741, 30

\bibitem[{Ackermann {et~al.}(2011{\natexlab{b}})Ackermann, Ajello, Allafort,
  Antolini, Atwood, Axelsson, Baldini, Ballet, Barbiellini, Bastieri, Bechtol,
  Bellazzini, Berenji, Blandford, Bloom, Bonamente, Borgland, Bottacini,
  Bouvier, Bregeon, Brigida, Bruel, Buehler, Burnett, Buson, Caliandro,
  Cameron, Caraveo, Casandjian, Cavazzuti, Cecchi, Charles, Cheung, Chiang,
  Ciprini, Claus, Cohen-Tanugi, Conrad, Costamante, Cutini, de~Angelis,
  de~Palma, Dermer, Digel, {do Couto e Silva}, Drell, Dubois, Escande, Favuzzi,
  Fegan, Ferrara, Finke, Focke, Fortin, Frailis, Fukazawa, Funk, Fusco,
  Gargano, Gasparrini, Gehrels, Germani, Giebels, Giglietto, Giommi, Giordano,
  Giroletti, Glanzman, Godfrey, Grenier, Grove, Guiriec, Gustafsson, Hadasch,
  Hayashida, Hays, Healey, Horan, Hou, Hughes, Iafrate, J\'{o}hannesson,
  Johnson, Johnson, Kamae, Katagiri, Kataoka, Kn\"{o}dlseder, Kuss, Lande,
  Larsson, Latronico, Longo, Loparco, Lott, Lovellette, Lubrano, Madejski,
  Mazziotta, McConville, McEnery, Michelson, Mitthumsiri, Mizuno, Moiseev,
  Monte, Monzani, Moretti, Morselli, Moskalenko, Murgia, Nakamori,
  Naumann-Godo, Nolan, Norris, Nuss, Ohno, Ohsugi, Okumura, Omodei, Orienti,
  Orlando, Ormes, Ozaki, Paneque, Parent, Pesce-Rollins, Pierbattista,
  Piranomonte, Piron, Pivato, Porter, Rain\`{o}, Rando, Razzano, Razzaque,
  Reimer, Reimer, Ritz, Rochester, Romani, Roth, Sanchez, Sbarra, Scargle,
  Schalk, Sgr\`{o}, Shaw, Siskind, Spandre, Spinelli, Strong, Suson, Tajima,
  Takahashi, Takahashi, Tanaka, Thayer, Thayer, Thompson, Tibaldo, Tinivella,
  Torres, Tosti, Troja, Uchiyama, Vandenbroucke, Vasileiou, Vianello, Vitale,
  Waite, Wallace, Wang, Winer, Wood, Wood, \& Zimmer}]{Ackermann2011b}
---. 2011{\natexlab{b}}, The Astrophysical Journal, 743, 171

\bibitem[{Ackermann {et~al.}(2012{\natexlab{a}})Ackermann, Ajello, Atwood,
  Baldini, Ballet, Barbiellini, Bastieri, Bechtol, Bellazzini, Berenji,
  Blandford, Bloom, Bonamente, Borgland, Brandt, Bregeon, Brigida, Bruel,
  Buehler, Buson, Caliandro, Cameron, Caraveo, Cavazzuti, Cecchi, Charles,
  Chekhtman, Chiang, Ciprini, Claus, Cohen-Tanugi, Conrad, Cutini, de~Angelis,
  de~Palma, Dermer, Digel, {do Couto e Silva}, Drell, Drlica-Wagner, Falletti,
  Favuzzi, Fegan, Ferrara, Focke, Fortin, Fukazawa, Funk, Fusco, Gaggero,
  Gargano, Germani, Giglietto, Giordano, Giroletti, Glanzman, Godfrey, Grove,
  Guiriec, Gustafsson, Hadasch, Hanabata, Harding, Hayashida, Hays, Horan, Hou,
  Hughes, J\'{o}hannesson, Johnson, Johnson, Kamae, Katagiri, Kataoka,
  Kn\"{o}dlseder, Kuss, Lande, Latronico, Lee, Lemoine-Goumard, Longo, Loparco,
  Lott, Lovellette, Lubrano, Mazziotta, McEnery, Michelson, Mitthumsiri,
  Mizuno, Monte, Monzani, Morselli, Moskalenko, Murgia, Naumann-Godo, Norris,
  Nuss, Ohsugi, Okumura, Omodei, Orlando, Ormes, Paneque, Panetta, Parent,
  Pesce-Rollins, Pierbattista, Piron, Pivato, Porter, Rain\`{o}, Rando,
  Razzano, Razzaque, Reimer, Reimer, Sadrozinski, Sgr\`{o}, Siskind, Spandre,
  Spinelli, Strong, Suson, Takahashi, Tanaka, Thayer, Thayer, Thompson,
  Tibaldo, Tinivella, Torres, Tosti, Troja, Usher, Vandenbroucke, Vasileiou,
  Vianello, Vitale, Waite, Wang, Winer, Wood, Wood, Yang, Ziegler, \&
  Zimmer}]{Ackermann2012a}
Ackermann, M., Ajello, M., Atwood, W.~B., {et~al.} 2012{\natexlab{a}}, The
  Astrophysical Journal, 750, 3

\bibitem[{Ackermann {et~al.}(2012{\natexlab{b}})Ackermann, Ajello, Ballet,
  Barbiellini, Bastieri, Belfiore, Bellazzini, Berenji, Blandford, Bloom,
  Bonamente, Borgland, Bregeon, Brigida, Bruel, Buehler, Buson, Caliandro,
  Cameron, Caraveo, Cavazzuti, Cecchi, \c{C}elik, Charles, Chaty, Chekhtman,
  Cheung, Chiang, Ciprini, Claus, Cohen-Tanugi, Corbel, Corbet, Cutini,
  de~Luca, den Hartog, de~Palma, Dermer, Digel, {do Couto e Silva}, Donato,
  Drell, Drlica-Wagner, Dubois, Dubus, Favuzzi, Fegan, Ferrara, Focke, Fortin,
  Fukazawa, Funk, Fusco, Gargano, Gasparrini, Gehrels, Germani, Giglietto,
  Giordano, Giroletti, Glanzman, Godfrey, Grenier, Grove, Guiriec, Hadasch,
  Hanabata, Harding, Hayashida, Hays, Hill, Hughes, J\'{o}hannesson, Johnson,
  Johnson, Kamae, Katagiri, Kataoka, Kerr, Kn\"{o}dlseder, Kuss, Lande, Longo,
  Loparco, Lovellette, Lubrano, Mazziotta, McEnery, Michelson, Mitthumsiri,
  Mizuno, Monte, Monzani, Morselli, Moskalenko, Murgia, Nakamori, Naumann-Godo,
  Norris, Nuss, Ohno, Ohsugi, Okumura, Omodei, Orlando, Ozaki, Paneque, Parent,
  Pesce-Rollins, Pierbattista, Piron, Pivato, Porter, Rain\`{o}, Rando,
  Razzano, Reimer, Reimer, Ritz, Romani, Roth, {Saz Parkinson}, Sgr\`{o},
  Siskind, Spandre, Spinelli, Suson, Takahashi, Tanaka, Thayer, Thayer,
  Thompson, Tibaldo, Tinivella, Torres, Tosti, Troja, Uchiyama, Usher,
  Vandenbroucke, Vianello, Vitale, Waite, Winer, Wood, Wood, Yang, Zimmer, Coe,
  {Di Mille}, Edwards, Filipovi\'{c}, Payne, Stevens, \&
  Torres}]{FermiLATCollaboration2012}
Ackermann, M., Ajello, M., Ballet, J., {et~al.} 2012{\natexlab{b}}, Science
  (New York, N.Y.), 335, 189

\bibitem[{Ackermann {et~al.}(2012{\natexlab{c}})Ackermann, Ajello, Albert,
  Allafort, Atwood, Axelsson, Baldini, Ballet, Barbiellini, Bastieri, Bechtol,
  Bellazzini, Bissaldi, Blandford, Bloom, Bogart, Bonamente, Borgland,
  Bottacini, Bouvier, Brandt, Bregeon, Brigida, Bruel, Buehler, Burnett, Buson,
  Caliandro, Cameron, Caraveo, Casandjian, Cavazzuti, Cecchi, \c{C}elik,
  Charles, Chaves, Chekhtman, Cheung, Chiang, Ciprini, Claus, Cohen-Tanugi,
  Conrad, Corbet, Cutini, D'Ammando, Davis, de~Angelis, DeKlotz, de~Palma,
  Dermer, Digel, {do Couto e Silva}, Drell, Drlica-Wagner, Dubois, Favuzzi,
  Fegan, Ferrara, Focke, Fortin, Fukazawa, Funk, Fusco, Gargano, Gasparrini,
  Gehrels, Giebels, Giglietto, Giordano, Giroletti, Glanzman, Godfrey, Grenier,
  Grove, Guiriec, Hadasch, Hayashida, Hays, Horan, Hou, Hughes, Jackson,
  Jogler, J\'{o}hannesson, Johnson, Johnson, Johnson, Kamae, Katagiri, Kataoka,
  Kerr, Kn\"{o}dlseder, Kuss, Lande, Larsson, Latronico, Lavalley,
  Lemoine-Goumard, Longo, Loparco, Lott, Lovellette, Lubrano, Mazziotta,
  McConville, McEnery, Mehault, Michelson, Mitthumsiri, Mizuno, Moiseev, Monte,
  Monzani, Morselli, Moskalenko, Murgia, Naumann-Godo, Nemmen, Nishino, Norris,
  Nuss, Ohno, Ohsugi, Okumura, Omodei, Orienti, Orlando, Ormes, Paneque,
  Panetta, Perkins, Pesce-Rollins, Pierbattista, Piron, Pivato, Porter,
  Racusin, Rain\`{o}, Rando, Razzano, Razzaque, Reimer, Reimer, Reposeur,
  Reyes, Ritz, Rochester, Romoli, Roth, Sadrozinski, Sanchez, {Saz Parkinson},
  Sbarra, Scargle, Sgr\`{o}, Siegal-Gaskins, Siskind, Spandre, Spinelli,
  Stephens, Suson, Tajima, Takahashi, Tanaka, Thayer, Thayer, Thompson,
  Tibaldo, Tinivella, Tosti, Troja, Usher, Vandenbroucke, {Van Klaveren},
  Vasileiou, Vianello, Vitale, Waite, Wallace, Winer, Wood, Wood, Wood, Yang,
  \& Zimmer}]{Ackermann2012}
Ackermann, M., Ajello, M., Albert, A., {et~al.} 2012{\natexlab{c}}, The
  Astrophysical Journal Supplement Series, 203, 4

\bibitem[{Aharonian {et~al.}(2005)Aharonian, Akhperjanian, Aye, Bazer-Bachi,
  Beilicke, Benbow, Berge, Berghaus, Bernl\"{o}hr, Boisson, Bolz, Braun,
  Breitling, Brown, {Bussons Gordo}, Chadwick, Chounet, Cornils, Costamante,
  Degrange, Djannati-Ata\"{\i}, Drury, Dubus, Emmanoulopoulos, Espigat,
  Feinstein, Fleury, Fontaine, Fuchs, Funk, Gallant, Giebels, Gillessen,
  Glicenstein, Goret, Hadjichristidis, Hauser, Heinzelmann, Henri, Hermann,
  Hinton, Hofmann, Holleran, Horns, de~Jager, Johnston, Kh\'{e}lifi, Kirk,
  Komin, Konopelko, Latham, {Le Gallou}, Lemi\`{e}re, Lemoine-Goumard, Leroy,
  Martineau-Huynh, Lohse, Marcowith, Masterson, McComb, de~Naurois, Nolan,
  Noutsos, Orford, Osborne, Ouchrif, Panter, Pelletier, Pita, P\"{u}hlhofer,
  Punch, Raubenheimer, Raue, Raux, Rayner, Redondo, Reimer, Reimer, Ripken,
  Rob, Rolland, Rowell, Sahakian, Saug\'{e}, Schlenker, Schlickeiser, Schuster,
  Schwanke, Siewert, Skj\ae~raasen, Sol, Steenkamp, Stegmann, Tavernet,
  Terrier, Th\'{e}oret, Tluczykont, Vasileiadis, Venter, Vincent, V\"{o}lk, \&
  Wagner}]{Aharonian2005}
Aharonian, F., Akhperjanian, A.~G., Aye, K.-M., {et~al.} 2005, Astronomy and
  Astrophysics, 442, 1

\bibitem[{Aharonian {et~al.}(2006)Aharonian, Akhperjanian, Bazer-Bachi,
  Beilicke, Benbow, Berge, Bernl\"{o}hr, Boisson, Bolz, Borrel, Braun, Brown,
  B\"{u}hler, B\"{u}sching, Carrigan, Chadwick, Chounet, Cornils, Costamante,
  Degrange, Dickinson, Djannati-Ata\"{\i}, {O'C. Drury}, Dubus, Egberts,
  Emmanoulopoulos, Espigat, Feinstein, Ferrero, Fiasson, Fontaine, Funk, Funk,
  F\"{u}\ss~ling, Gallant, Giebels, Glicenstein, Goret, Hadjichristidis,
  Hauser, Hauser, Heinzelmann, Henri, Hermann, Hinton, Hoffmann, Hofmann,
  Holleran, Horns, Jacholkowska, de~Jager, Kendziorra, Kh\'{e}lifi, Komin,
  Konopelko, Kosack, Latham, {Le Gallou}, Lemi\`{e}re, Lemoine-Goumard, Lohse,
  Martin, Martineau-Huynh, Marcowith, Masterson, Maurin, McComb, Moulin,
  de~Naurois, Nedbal, Nolan, Noutsos, Orford, Osborne, Ouchrif, Panter,
  Pelletier, Pita, P\"{u}hlhofer, Punch, Raubenheimer, Raue, Rayner, Reimer,
  Reimer, Ripken, Rob, Rolland, Rowell, Sahakian, Santangelo, Saug\'{e},
  Schlenker, Schlickeiser, Schr\"{o}der, Schwanke, Schwarzburg, Shalchi, Sol,
  Spangler, Spanier, Steenkamp, Stegmann, Superina, Tavernet, Terrier,
  Tluczykont, van Eldik, Vasileiadis, Venter, Vincent, V\"{o}lk, Wagner, \&
  Ward}]{Aharonian2006}
Aharonian, F., Akhperjanian, A.~G., Bazer-Bachi, A.~R., {et~al.} 2006,
  Astronomy and Astrophysics, 460, 743

\bibitem[{Aharonian {et~al.}(2008)Aharonian, Akhperjanian, {Barres de Almeida},
  Bazer-Bachi, Becherini, Behera, Benbow, Bernl\"{o}hr, Boisson, Bochow,
  Borrel, Braun, Brion, Brucker, Brun, B\"{u}hler, Bulik, B\"{u}sching,
  Boutelier, Carrigan, Chadwick, Charbonnier, Chaves, Cheesebrough, Chounet,
  Clapson, Coignet, Dalton, Degrange, Deil, Dickinson, Djannati-Ata\"{\i},
  Domainko, {O'C. Drury}, Dubois, Dubus, Dyks, Dyrda, Egberts, Emmanoulopoulos,
  Espigat, Farnier, Feinstein, Fiasson, F\"{o}rster, Fontaine, F\"{u}\ss~ling,
  Gabici, Gallant, G\'{e}rard, Giebels, Glicenstein, Gl\"{u}ck, Goret,
  Hadjichristidis, Hauser, Hauser, Heinz, Heinzelmann, Henri, Hermann, Hinton,
  Hoffmann, Hofmann, Holleran, Hoppe, Horns, Jacholkowska, de~Jager, Jung,
  Katarzyński, Kaufmann, Kendziorra, Kerschhaggl, Khangulyan, Kh\'{e}lifi,
  Keogh, Komin, Kosack, Lamanna, Lenain, Lohse, Marandon, Martin,
  Martineau-Huynh, Marcowith, Maurin, McComb, Medina, Moderski, Moulin,
  Naumann-Godo, de~Naurois, Nedbal, Nekrassov, Niemiec, Nolan, Ohm, Olive, {de
  O\~{n}a Wilhelmi}, Orford, Osborne, Ostrowski, Panter, Pedaletti, Pelletier,
  Petrucci, Pita, P\"{u}hlhofer, Punch, Quirrenbach, Raubenheimer, Raue,
  Rayner, Renaud, Rieger, Ripken, Rob, Rosier-Lees, Rowell, Rudak, Rulten,
  Ruppel, Sahakian, Santangelo, Schlickeiser, Sch\"{o}ck, Schr\"{o}der,
  Schwanke, Schwarzburg, Schwemmer, Shalchi, Skilton, Sol, Spangler, Stawarz,
  Steenkamp, Stegmann, Superina, Tam, Tavernet, Terrier, Tibolla, van Eldik,
  Vasileiadis, Venter, Vialle, Vincent, Vivier, V\"{o}lk, Volpe, Wagner, Ward,
  Zdziarski, \& Zech}]{Aharonian2008}
Aharonian, F., Akhperjanian, A.~G., {Barres de Almeida}, U., {et~al.} 2008,
  Astronomy and Astrophysics, 492, L25

\bibitem[{Aharonian {et~al.}(2009)Aharonian, Akhperjanian, Anton, {Barres de
  Almeida}, Bazer-Bachi, Becherini, Behera, Benbow, Bernl\"{o}hr, Boisson,
  Bochow, Borrel, Brion, Brucker, Brun, B\"{u}hler, Bulik, B\"{u}sching,
  Boutelier, Chadwick, Charbonnier, Chaves, Cheesebrough, Chounet, Clapson,
  Coignet, Costamante, Dalton, Daniel, Davids, Degrange, Deil, Dickinson,
  Djannati-Ata\"{\i}, Domainko, Drury, Dubois, Dubus, Dyks, Dyrda, Egberts,
  Emmanoulopoulos, Espigat, Farnier, Feinstein, Fiasson, F\"{o}rster, Fontaine,
  F\"{u}\ss~ling, Gabici, Gallant, G\'{e}rard, Giebels, Glicenstein, Gl\"{u}ck,
  Goret, G\"{o}hring, Hauser, Hauser, Heinz, Heinzelmann, Henri, Hermann,
  Hinton, Hoffmann, Hofmann, Holleran, Hoppe, Horns, Jacholkowska, de~Jager,
  Jahn, Jung, Katarzyński, Katz, Kaufmann, Kendziorra, Kerschhaggl,
  Khangulyan, Kh\'{e}lifi, Keogh, Klu\'{z}niak, Kneiske, Komin, Kosack,
  Lamanna, Lenain, Lohse, Marandon, Martin, Martineau-Huynh, Marcowith, Maurin,
  McComb, Medina, Moderski, Monard, Moulin, Naumann-Godo, de~Naurois, Nedbal,
  Nekrassov, Niemiec, Nolan, Ohm, Olive, {de O\~{n}a Wilhelmi}, Orford,
  Ostrowski, Panter, {Paz Arribas}, Pedaletti, Pelletier, Petrucci, Pita,
  P\"{u}hlhofer, Punch, Quirrenbach, Raubenheimer, Raue, Rayner, Renaud,
  Rieger, Ripken, Rob, Rosier-Lees, Rowell, Rudak, Rulten, Ruppel, Sahakian,
  Santangelo, Schlickeiser, Sch\"{o}ck, Schr\"{o}der, Schwanke, Schwarzburg,
  Schwemmer, Shalchi, Sikora, Skilton, Sol, Spangler, Stawarz, Steenkamp,
  Stegmann, Superina, Szostek, Tam, Tavernet, Terrier, Tibolla, Tluczykont, van
  Eldik, Vasileiadis, Venter, Venter, Vialle, Vincent, Vivier, V\"{o}lk, Volpe,
  Wagner, Ward, Zdziarski, \& Zech}]{Aharonian2009}
Aharonian, F., Akhperjanian, A.~G., Anton, G., {et~al.} 2009, Astronomy and
  Astrophysics, 502, 749

\bibitem[{Albert {et~al.}(2006)Albert, Aliu, Anderhub, Antoranz, Armada,
  Asensio, Baixeras, Barrio, Bartelt, Bartko, Bastieri, Bavikadi, Bednarek,
  Berger, Bigongiari, Biland, Bisesi, Bock, Bordas, Bosch-Ramon, Bretz,
  Britvitch, Camara, Carmona, Chilingarian, Ciprini, Coarasa, Commichau,
  Contreras, Cortina, Curtef, Danielyan, Dazzi, {De Angelis}, {de Los Reyes},
  {De Lotto}, Domingo-Santamar\'{\i}a, Dorner, Doro, Errando, Fagiolini,
  Ferenc, Fern\'{a}ndez, Firpo, Flix, Fonseca, Font, Fuchs, Galante,
  Garczarczyk, Gaug, Giller, Goebel, Hakobyan, Hayashida, Hengstebeck,
  H\"{o}hne, Hose, Hsu, Isar, Jacon, Kalekin, Kosyra, Kranich, Laatiaoui,
  Laille, Lenisa, Liebing, Lindfors, Lombardi, Longo, L\'{o}pez, L\'{o}pez,
  Lorenz, Lucarelli, Majumdar, Maneva, Mannheim, Mansutti, Mariotti,
  Mart\'{\i}nez, Mase, Mazin, Merck, Meucci, Meyer, Miranda, Mirzoyan,
  Mizobuchi, Moralejo, Nilsson, O\~{n}a Wilhelmi, Ordu\~{n}a, Otte, Oya,
  Paneque, Paoletti, Paredes, Pasanen, Pascoli, Pauss, Pavel, Pegna, Persic,
  Peruzzo, Piccioli, Poller, Pooley, Prandini, Raymers, Rhode, Rib\'{o}, Rico,
  Riegel, Rissi, Robert, Romero, R\"{u}gamer, Saggion, S\'{a}nchez, Sartori,
  Scalzotto, Scapin, Schmitt, Schweizer, Shayduk, Shinozaki, Shore, Sidro,
  Sillanp\"{a}\"{a}, Sobczynska, Stamerra, Stark, Takalo, Temnikov, Tescaro,
  Teshima, Tonello, Torres, Torres, Turini, Vankov, Vitale, Wagner, Wibig,
  Wittek, Zanin, \& Zapatero}]{Albert2006}
Albert, J., Aliu, E., Anderhub, H., {et~al.} 2006, Science (New York, N.Y.),
  312, 1771

\bibitem[{Atwood {et~al.}(2009)Atwood, Abdo, Ackermann, Althouse, Anderson,
  Axelsson, Baldini, Ballet, Band, Barbiellini, Bartelt, Bastieri, Baughman,
  Bechtol, B\'{e}d\'{e}r\`{e}de, Bellardi, Bellazzini, Berenji, Bignami,
  Bisello, Bissaldi, Blandford, Bloom, Bogart, Bonamente, Bonnell, Borgland,
  Bouvier, Bregeon, Brez, Brigida, Bruel, Burnett, Busetto, Caliandro, Cameron,
  Caraveo, Carius, Carlson, Casandjian, Cavazzuti, Ceccanti, Cecchi, Charles,
  Chekhtman, Cheung, Chiang, Chipaux, Cillis, Ciprini, Claus, Cohen-Tanugi,
  Condamoor, Conrad, Corbet, Corucci, Costamante, Cutini, Davis, Decotigny,
  DeKlotz, Dermer, de~Angelis, Digel, {do Couto e Silva}, Drell, Dubois,
  Dumora, Edmonds, Fabiani, Farnier, Favuzzi, Flath, Fleury, Focke, Funk,
  Fusco, Gargano, Gasparrini, Gehrels, Gentit, Germani, Giebels, Giglietto,
  Giommi, Giordano, Glanzman, Godfrey, Grenier, Grondin, Grove, Guillemot,
  Guiriec, Haller, Harding, Hart, Hays, Healey, Hirayama, Hjalmarsdotter, Horn,
  Hughes, J\'{o}hannesson, Johansson, Johnson, Johnson, Johnson, Johnson,
  Kamae, Katagiri, Kataoka, Kavelaars, Kawai, Kelly, Kerr, Klamra,
  Kn\"{o}dlseder, Kocian, Komin, Kuehn, Kuss, Landriu, Latronico, Lee, Lee,
  Lemoine-Goumard, Lionetto, Longo, Loparco, Lott, Lovellette, Lubrano,
  Madejski, Makeev, Marangelli, Massai, Mazziotta, McEnery, Menon, Meurer,
  Michelson, Minuti, Mirizzi, Mitthumsiri, Mizuno, Moiseev, Monte, Monzani,
  Moretti, Morselli, Moskalenko, Murgia, Nakamori, Nishino, Nolan, Norris,
  Nuss, Ohno, Ohsugi, Omodei, Orlando, Ormes, Paccagnella, Paneque, Panetta,
  Parent, Pearce, Pepe, Perazzo, Pesce-Rollins, Picozza, Pieri, Pinchera,
  Piron, Porter, Poupard, Rain\`{o}, Rando, Rapposelli, Razzano, Reimer,
  Reimer, Reposeur, Reyes, Ritz, Rochester, Rodriguez, Romani, Roth, Russell,
  Ryde, Sabatini, Sadrozinski, Sanchez, Sander, Sapozhnikov, Parkinson,
  Scargle, Schalk, Scolieri, Sgr\`{o}, Share, Shaw, Shimokawabe, Shrader,
  Sierpowska-Bartosik, Siskind, Smith, Smith, Spandre, Spinelli, Starck,
  Stephens, Strickman, Strong, Suson, Tajima, Takahashi, Takahashi, Tanaka,
  Tenze, Tether, Thayer, Thayer, Thompson, Tibaldo, Tibolla, Torres, Tosti,
  Tramacere, Turri, Usher, Vilchez, Vitale, Wang, Watters, Winer, Wood, Ylinen,
  \& Ziegler}]{Atwood2009}
Atwood, W.~B., Abdo, A.~A., Ackermann, M., {et~al.} 2009, The Astrophysical
  Journal, 697, 1071

\bibitem[{Buehler {et~al.}(2012)Buehler, Scargle, Blandford, Baldini, Baring,
  Belfiore, Charles, Chiang, D'Ammando, Dermer, Funk, Grove, Harding, Hays,
  Kerr, Massaro, Mazziotta, Romani, {Saz Parkinson}, Tennant, \&
  Weisskopf}]{Buehler2012}
Buehler, R., Scargle, J.~D., Blandford, R.~D., {et~al.} 2012, The Astrophysical
  Journal, 749, 26

\bibitem[{Cheung {et~al.}(2012{\natexlab{a}})Cheung, Donato, Gehrels,
  Sokolovsky, \& Giroletti}]{Cheung2012}
Cheung, C.~C., Donato, D., Gehrels, N., Sokolovsky, K.~V., \& Giroletti, M.
  2012{\natexlab{a}}, The Astrophysical Journal, 756, 33

\bibitem[{Cheung {et~al.}(2012{\natexlab{b}})Cheung, Glanzman, \&
  Hill}]{Cheung2012a}
Cheung, C.~C., Glanzman, T., \& Hill, A.~B. 2012{\natexlab{b}}, The
  Astronomer's Telegram, 4284

\bibitem[{Cheung {et~al.}(2012{\natexlab{c}})Cheung, Hays, Venters, Donato, \&
  Corbet}]{Cheung2012b}
Cheung, C.~C., Hays, E., Venters, T., Donato, D., \& Corbet, R. H.~D.
  2012{\natexlab{c}}, The Astronomer's Telegram, 4224

\bibitem[{Cheung {et~al.}(2012{\natexlab{d}})Cheung, Shore, {De Gennaro
  Aquino}, Charbonnel, Edlin, Hays, Corbet, \& Wood}]{CheungC.C.2012}
Cheung, C.~C., Shore, S.~N., {De Gennaro Aquino}, I., {et~al.}
  2012{\natexlab{d}}, The Astronomer's Telegram, 4310

\bibitem[{Chomiuk {et~al.}(2013)Chomiuk, Strader, Landt, \&
  Cheung}]{Chomiuk2013}
Chomiuk, L., Strader, J., Landt, H., \& Cheung, C.~C. 2013, The Astronomer's
  Telegram, 4777

\bibitem[{Ciprini {et~al.}(2012)Ciprini, Hays, \& Cheung}]{Ciprini2012}
Ciprini, S., Hays, E., \& Cheung, C.~C. 2012, The Astronomer's Telegram, 3978

\bibitem[{Ciprini \& Hays(2012)}]{CipriniStefano2012}
Ciprini, S., \& Hays, E.~A. 2012, The Astronomer's Telegram, 4182

\bibitem[{Donato \& Perkins(2011)}]{Donato2011}
Donato, D., \& Perkins, J.~S. 2011, The Astronomer's Telegram, 3452

\bibitem[{Dubus(2007)}]{DubusG.2007}
Dubus, G. 2007, {Gamma-ray emission from binaries with GLAST}, ed. C.~C.
  {Bouvier J., Chalabaev A.}, 163

\bibitem[{{Fermi-LAT Collaboration}(2013)}]{Fermi-LATCollaboration}
{Fermi-LAT Collaboration}. 2013, https://www-glast.stanford.edu/pub\_data/585

\bibitem[{Foschini {et~al.}(2012)Foschini, Angelakis, Fuhrmann, Ghisellini,
  Hovatta, Lahteenmaki, Lister, Braito, Gallo, Hamilton, Kino, Komossa,
  Pushkarev, Thompson, Tibolla, Tramacere, Carrami\~{n}ana, Carrasco, Falcone,
  Giroletti, Grupe, Kovalev, Krichbaum, Max-Moerbeck, Nestoras, Pearson,
  Porras, Readhead, Recillas, Richards, Riquelme, Sievers, Tammi, Tornikoski,
  Ungerechts, Zensus, Celotti, Bonnoli, Doi, Maraschi, Tagliaferri, \&
  Tavecchio}]{Foschini2012}
Foschini, L., Angelakis, E., Fuhrmann, L., {et~al.} 2012, Astronomy \&
  Astrophysics, 548, A106

\bibitem[{Hadasch {et~al.}(2012)Hadasch, Torres, Tanaka, Corbet, Hill, Dubois,
  Dubus, Glanzman, Corbel, Li, Chen, Zhang, Caliandro, Kerr, Richards,
  Max-Moerbeck, Readhead, \& Pooley}]{Hadasch2012}
Hadasch, D., Torres, D.~F., Tanaka, T., {et~al.} 2012, The Astrophysical
  Journal, 749, 54

\bibitem[{Hinton {et~al.}(2009)Hinton, Skilton, Funk, Brucker, Aharonian,
  Dubus, Fiasson, Gallant, Hofmann, Marcowith, \& Reimer}]{Hinton2009}
Hinton, J.~A., Skilton, J.~L., Funk, S., {et~al.} 2009, The Astrophysical
  Journal, 690, L101

\bibitem[{Kataoka {et~al.}(2010)Kataoka, Stawarz, Cheung, Tosti, Cavazzuti,
  Celotti, Nishino, Fukazawa, Thompson, \& McConville}]{Kataoka2010}
Kataoka, J., Stawarz, Å., Cheung, C.~C., {et~al.} 2010, The Astrophysical
  Journal, 715, 554

\bibitem[{Morh\'{a}\v{c} {et~al.}(2000)Morh\'{a}\v{c}, Kliman, Matou\v{s}ek,
  Veselsk\'{y}, \& Turzo}]{Morhac2000}
Morh\'{a}\v{c}, M., Kliman, J., Matou\v{s}ek, V., Veselsk\'{y}, M., \& Turzo,
  I. 2000, Nuclear Instruments and Methods in Physics Research Section A:
  Accelerators, Spectrometers, Detectors and Associated Equipment, 443, 108

\bibitem[{Nea {et~al.}(2001)Nea, Milkova, \& Nea}]{Nea2001}
Nea, J., Milkova, E., \& Nea, H. 2001, Discrete Mathematics, 233, 3

\bibitem[{Neronov {et~al.}(2012)Neronov, Malyshev, Chernyakova, \&
  Lutovinov}]{Neronov2012}
Neronov, A., Malyshev, D., Chernyakova, M., \& Lutovinov, A. 2012, Astronomy \&
  Astrophysics, 543, L9

\bibitem[{Nolan {et~al.}(2012)Nolan, Abdo, Ackermann, Ajello, Allafort,
  Antolini, Atwood, Axelsson, Baldini, Ballet, Barbiellini, Bastieri, Bechtol,
  Belfiore, Bellazzini, Berenji, Bignami, Blandford, Bloom, Bonamente, Bonnell,
  Borgland, Bottacini, Bouvier, Brandt, Bregeon, Brigida, Bruel, Buehler,
  Burnett, Buson, Caliandro, Cameron, Campana, Ca\~{n}adas, Cannon, Caraveo,
  Casandjian, Cavazzuti, Ceccanti, Cecchi, \c{C}elik, Charles, Chekhtman,
  Cheung, Chiang, Chipaux, Ciprini, Claus, Cohen-Tanugi, Cominsky, Conrad,
  Corbet, Cutini, D'Ammando, Davis, de~Angelis, DeCesar, DeKlotz, {De Luca},
  den Hartog, de~Palma, Dermer, Digel, {do Couto e Silva}, Drell,
  Drlica-Wagner, Dubois, Dumora, Enoto, Escande, Fabiani, Falletti, Favuzzi,
  Fegan, Ferrara, Focke, Fortin, Frailis, Fukazawa, Funk, Fusco, Gargano,
  Gasparrini, Gehrels, Germani, Giebels, Giglietto, Giommi, Giordano,
  Giroletti, Glanzman, Godfrey, Grenier, Grondin, Grove, Guillemot, Guiriec,
  Gustafsson, Hadasch, Hanabata, Harding, Hayashida, Hays, Hill, Horan, Hou,
  Hughes, Iafrate, Itoh, J\'{o}hannesson, Johnson, Johnson, Johnson, Johnson,
  Kamae, Katagiri, Kataoka, Katsuta, Kawai, Kerr, Kn\"{o}dlseder, Kocevski,
  Kuss, Lande, Landriu, Latronico, Lemoine-Goumard, Lionetto, {Llena Garde},
  Longo, Loparco, Lott, Lovellette, Lubrano, Madejski, Marelli, Massaro,
  Mazziotta, McConville, McEnery, Mehault, Michelson, Minuti, Mitthumsiri,
  Mizuno, Moiseev, Mongelli, Monte, Monzani, Morselli, Moskalenko, Murgia,
  Nakamori, Naumann-Godo, Norris, Nuss, Nymark, Ohno, Ohsugi, Okumura, Omodei,
  Orlando, Ormes, Ozaki, Paneque, Panetta, Parent, Perkins, Pesce-Rollins,
  Pierbattista, Pinchera, Piron, Pivato, Porter, Racusin, Rain\`{o}, Rando,
  Razzano, Razzaque, Reimer, Reimer, Reposeur, Ritz, Rochester, Romani, Roth,
  Rousseau, Ryde, Sadrozinski, Salvetti, Sanchez, {Saz Parkinson}, Sbarra,
  Scargle, Schalk, Sgr\`{o}, Shaw, Shrader, Siskind, Smith, Spandre, Spinelli,
  Stephens, Strickman, Suson, Tajima, Takahashi, Takahashi, Tanaka, Thayer,
  Thayer, Thompson, Tibaldo, Tibolla, Tinebra, Tinivella, Torres, Tosti, Troja,
  Uchiyama, Vandenbroucke, {Van Etten}, {Van Klaveren}, Vasileiou, Vianello,
  Vitale, Waite, Wallace, Wang, Werner, Winer, Wood, Wood, Wood, Yang, \&
  Zimmer}]{Nolan2012}
Nolan, P.~L., Abdo, A.~A., Ackermann, M., {et~al.} 2012, The Astrophysical
  Journal Supplement Series, 199, 31

\bibitem[{Orienti \& D'Ammando(2012)}]{Orienti2012}
Orienti, M., \& D'Ammando, F. 2012, The Astronomer's Telegram, 3999

\bibitem[{Percy(2007)}]{Percy2007}
Percy, J.~R. 2007, {Understanding Variable Stars (Cambridge Astrophysics)}
  (Cambridge University Press), 374

\bibitem[{Reitberger {et~al.}(2012)Reitberger, Reimer, Reimer, Werner, Egberts,
  \& Takahashi}]{Reitberger2012}
Reitberger, K., Reimer, O., Reimer, A., {et~al.} 2012, Astronomy \&
  Astrophysics, 544, A98

\bibitem[{Sabatini {et~al.}(2010)Sabatini, Tavani, Striani, Bulgarelli,
  Vittorini, Piano, {Del Monte}, Feroci, de~Pasquale, Trifoglio, Gianotti,
  Argan, Barbiellini, Caraveo, Cattaneo, Chen, D'Ammando, Costa, {De Paris},
  {Di Cocco}, Donnarumma, Evangelista, Ferrari, Fiorini, Fuschino, Galli,
  Giuliani, Giusti, Labanti, Lazzarotto, Lipari, Longo, Marisaldi, Mereghetti,
  Morelli, Moretti, Morselli, Pacciani, Pellizzoni, Perotti, Picozza, Pilia,
  Pucella, Prest, Rapisarda, Rappoldi, Rubini, Scalise, Soffitta, Trois,
  Vallazza, Vercellone, Zambra, Zanello, Pittori, Verrecchia, Santolamazza,
  Giommi, Colafrancesco, Antonelli, \& Salotti}]{Sabatini2010}
Sabatini, S., Tavani, M., Striani, E., {et~al.} 2010, The Astrophysical
  Journal, 712, L10

\bibitem[{Tavani {et~al.}(2009)Tavani, Sabatini, Pian, Bulgarelli, Caraveo,
  Viotti, Corcoran, Giuliani, Pittori, Verrecchia, Vercellone, Mereghetti,
  Argan, Barbiellini, Boffelli, Cattaneo, Chen, Cocco, D'Ammando, Costa, {De
  Paris}, Monte, Cocco, Donnarumma, Evangelista, Ferrari, Feroci, Fiorini,
  Froysland, Fuschino, Galli, Gianotti, Labanti, Lapshov, Lazzarotto, Lipari,
  Longo, Marisaldi, Mastropietro, Morelli, Moretti, Morselli, Pacciani,
  Pellizzoni, Perotti, Piano, Picozza, Pilia, Porrovecchio, Pucella, Prest,
  Rapisarda, Rappoldi, Rubini, Soffitta, Trifoglio, Trois, Vallazza, Vittorini,
  Zambra, Zanello, Santolamazza, Giommi, Colafrancesco, Antonelli, \&
  Salotti}]{Tavani2009}
Tavani, M., Sabatini, S., Pian, E., {et~al.} 2009, The Astrophysical Journal,
  698, L142

\bibitem[{Tavani {et~al.}(2011)Tavani, Bulgarelli, Vittorini, Pellizzoni,
  Striani, Caraveo, Weisskopf, Tennant, Pucella, Trois, Costa, Evangelista,
  Pittori, Verrecchia, {Del Monte}, Campana, Pilia, {De Luca}, Donnarumma,
  Horns, Ferrigno, Heinke, Trifoglio, Gianotti, Vercellone, Argan, Barbiellini,
  Cattaneo, Chen, Contessi, D'Ammando, DePris, {Di Cocco}, {Di Persio}, Feroci,
  Ferrari, Galli, Giuliani, Giusti, Labanti, Lapshov, Lazzarotto, Lipari,
  Longo, Fuschino, Marisaldi, Mereghetti, Morelli, Moretti, Morselli, Pacciani,
  Perotti, Piano, Picozza, Prest, Rapisarda, Rappoldi, Rubini, Sabatini,
  Soffitta, Vallazza, Zambra, Zanello, Lucarelli, Santolamazza, Giommi,
  Salotti, \& Bignami}]{Tavani2011}
Tavani, M., Bulgarelli, A., Vittorini, V., {et~al.} 2011, Science (New York,
  N.Y.), 331, 736

\bibitem[{Vandenbroucke {et~al.}(2010)Vandenbroucke, Buehler, Ajello, Bechtol,
  Bellini, Bolte, Cheung, Civano, Donato, Fuhrmann, Funk, Healey, Hill, Knigge,
  Madejski, Romani, Santander-Garc\'{\i}a, Shaw, Steeghs, Torres, {Van Etten},
  \& Williams}]{Vandenbroucke2010}
Vandenbroucke, J., Buehler, R., Ajello, M., {et~al.} 2010, The Astrophysical
  Journal, 718, L166

\end{thebibliography}
\clearpage
\begin{landscape}

\begin{deluxetable}{lrrrrrrrrrrr} 
\tablecolumns{8} 
\tablewidth{0pc} 
\tablecaption{List of FAVA sources. The first column shows the FAVA identification number (ID), which is composed of the right ascension and declination of the source (J2000). The following columns show the Galactic coordinates and the statistical position error at 68\% confidence $R_{68}$. The systematic error on the source position is 0.1$^\circ$. Also shown are the total number of detected flares N$^{f}$, number of flares with detections at high energy N$^{f}_{he}$, and number of flares corresponding to negative flux variations N$^{f}_{neg}$. The last columns show the associated \emph{Fermi}-LAT source and the counterpart at longer wavelength found within a distance $R_s$ (see text). For sources first announced via ATels we post the telegrams number. $R_s$ is derived as the 99\% statistical error plus the systematic error.} 
\tablehead{ 
\colhead{1FAV ID} & \colhead{$l$ [$^{\circ}$]}   & \colhead{$b$ [$^{\circ}$]}& \colhead{$R_{68}$ [$^{\circ}$]}     & \colhead{N$^{f}$}    & \colhead{N$^{f}_{he}$} & \colhead{N$^{f}_{neg}$} & \colhead{$R_s$ [$^{\circ}$]}    & \colhead{LAT Assoc.} & \colhead{ATel} & \colhead{Assoc.} } 
\startdata
J0019$-$05 & 102.0 & $-$66.9 & 0.8 & 1 & 0 & 0 & 1.8 & 2FGL J0017.6$-$0510 & \nodata & PMN J0017$-$0512 &\\
J0029$-$55 & 309.7 & $-$61.8 & 0.8 & 1 & 0 & 0 & 1.8 & 2FGL J0032.7$-$5521 & \nodata & \nodata &\\
J0031$-$02 & 111.1 & $-$64.7 & 0.6 & 1 & 1 & 0 & 1.4 & \nodata & \nodata & \nodata &\\
J0104+58 & 124.6 & $-$4.5 & 0.3 & 7 & 5 & 0 & 0.7 & 2FGL J0102.7+5827 & \nodata & TXS 0059+581 &\\
J0112+23 & 129.1 & $-$39.6 & 0.6 & 1 & 1 & 0 & 1.4 & 2FGL J0112.1+2245 & \nodata & S2 0109+22 &\\
J0112+61 & 125.5 & $-$1.5 & 0.3 & 7 & 3 & 0 & 0.8 & 2FGL J0109.9+6132 & \nodata & TXS 0106+612 &\\
J0113+32 & 128.4 & $-$30.4 & 0.4 & 4 & 2 & 0 & 1.0 & 2FGL J0112.8+3208 & \nodata & 4C 31.03 &\\
J0115$-$11 & 144.2 & $-$73.4 & 0.3 & 5 & 5 & 0 & 0.7 & 2FGL J0116.0$-$1134 & \nodata & PKS 0113$-$118 &\\
J0139+47 & 131.2 & $-$14.6 & 0.8 & 1 & 0 & 0 & 1.8 & 2FGL J0136.9+4751 & \nodata & OC 457 &\\
J0203$-$17 & 186.2 & $-$70.6 & 0.8 & 1 & 0 & 0 & 1.8 & 2FGL J0205.3$-$1657 & \nodata & PKS 0202$-$17 &\\
J0203+15 & 147.5 & $-$43.9 & 0.6 & 1 & 1 & 0 & 1.4 & 2FGL J0205.0+1514 & \nodata & 4C +15.05 &\\
J0203+30 & 140.9 & $-$30.0 & 0.3 & 5 & 5 & 0 & 0.7 & \nodata & \nodata & \nodata &\\
J0209$-$51 & 276.6 & $-$61.8 & 0.3 & 10 & 5 & 0 & 0.7 & 2FGL J0210.7$-$5102 & \nodata & PKS 0208$-$512 &\\
J0212+10 & 153.6 & $-$47.9 & 0.3 & 5 & 5 & 0 & 0.7 & \nodata & \nodata & \nodata &\\
J0215+01 & 161.2 & $-$54.7 & 0.6 & 1 & 1 & 0 & 1.4 & 2FGL J0217.9+0143 & \nodata & PKS 0215+015 &\\
J0219$-$27 & 220.6 & $-$70.4 & 0.6 & 2 & 1 & 0 & 1.4 & \nodata & \nodata & \nodata &\\
J0219+35 & 142.3 & $-$23.8 & 0.6 & 1 & 1 & 0 & 1.4 & 2FGL J0221.0+3555 & \nodata & S4 0218+35 &\\
J0225+42 & 140.6 & $-$16.7 & 0.3 & 4 & 3 & 0 & 0.8 & 2FGL J0222.6+4302 & \nodata & 3C 66A &\\
J0227$-$56 & 279.1 & $-$56.0 & 0.6 & 1 & 1 & 0 & 1.4 & \nodata & \nodata & \nodata &\\
J0228$-$36 & 243.5 & $-$67.4 & 0.6 & 3 & 1 & 0 & 1.4 & 2FGL J0229.3$-$3644 & \nodata & PKS 0227$-$369 &\\
J0234$-$61 & 283.5 & $-$51.4 & 0.3 & 3 & 3 & 0 & 0.8 & 2FGL J0237.1$-$6136 & \nodata & PKS 0235$-$618 &\\
J0237+60 & 135.5 & 0.4 & 0.4 & 4 & 0 & 2 & 0.9 & 2FGL J0240.5+6113 & \nodata & LS I+61 303 &\\
J0238+16 & 156.9 & $-$39.2 & 0.1 & 16 & 14 & 1 & 0.4 & 2FGL J0238.7+1637 & \nodata & AO 0235+164 &\\
J0238+28 & 149.6 & $-$28.7 & 0.2 & 13 & 7 & 0 & 0.6 & 2FGL J0237.8+2846 & \nodata & 4C +28.07 &\\
J0246$-$46 & 261.4 & $-$60.1 & 0.4 & 3 & 2 & 0 & 1.0 & 2FGL J0245.9$-$4652 & \nodata & PKS 0244$-$470 &\\
J0251$-$22 & 210.2 & $-$62.5 & 0.3 & 7 & 4 & 0 & 0.7 & 2FGL J0252.7$-$2218 & \nodata & PKS 0250$-$225 &\\
J0302$-$24 & 215.2 & $-$60.4 & 0.3 & 3 & 3 & 0 & 0.8 & 2FGL J0303.4$-$2407 & \nodata & PKS 0301$-$243 &\\
J0308+10 & 169.2 & $-$39.8 & 0.3 & 6 & 4 & 0 & 0.7 & 2FGL J0309.1+1027 & \nodata & PKS 0306+102 &\\
J0312+01 & 178.7 & $-$45.7 & 0.8 & 1 & 0 & 0 & 1.8 & 2FGL J0312.6+0132 & \nodata & PKS 0310+013 &\\
J0320+41 & 150.6 & $-$13.1 & 0.3 & 3 & 3 & 0 & 0.8 & 2FGL J0319.8+4130 & \nodata & NGC 1275 &\\
J0329+22 & 164.6 & $-$27.4 & 0.8 & 1 & 0 & 0 & 1.8 & 2FGL J0326.1+2226 & \nodata & TXS 0322+222 &\\
J0333$-$40 & 245.2 & $-$54.1 & 0.6 & 1 & 1 & 0 & 1.4 & 2FGL J0334.2$-$4008 & \nodata & PKS 0332$-$403 &\\
J0333+32 & 158.4 & $-$18.9 & 0.8 & 1 & 0 & 0 & 1.8 & \nodata & \nodata & \nodata &\\
J0341$-$02 & 189.0 & $-$42.5 & 0.8 & 1 & 0 & 0 & 1.8 & 2FGL J0339.4$-$0144 & \nodata & PKS 0336$-$01 &\\
J0342$-$25 & 220.0 & $-$51.7 & 0.6 & 1 & 1 & 0 & 1.4 & \nodata & \nodata & \nodata &\\
J0350+79 & 130.5 & 19.8 & 0.6 & 1 & 1 & 0 & 1.4 & 2FGL J0354.1+8010 & \nodata & S5 0346+80 &\\
J0351$-$21 & 214.5 & $-$48.8 & 0.8 & 1 & 0 & 0 & 1.8 & 2FGL J0350.0$-$2104 & \nodata & PKS 0347$-$211 &\\
J0402$-$36 & 237.9 & $-$48.8 & 0.2 & 14 & 7 & 1 & 0.6 & 2FGL J0403.9$-$3604 & \nodata & PKS 0402$-$362 &\\
J0422$-$01 & 195.3 & $-$33.4 & 0.6 & 1 & 1 & 0 & 1.4 & 2FGL J0423.2$-$0120 & \nodata & PKS 0420$-$01 &\\
J0427$-$60 & 271.0 & $-$40.9 & 0.6 & 1 & 1 & 0 & 1.4 & 2FGL J0433.4$-$6029 & \nodata & PKS 0432$-$606 &\\
J0428$-$38 & 240.9 & $-$43.7 & 0.2 & 30 & 11 & 7 & 0.5 & 2FGL J0428.6$-$3756 & \nodata & PKS 0426$-$380 &\\
J0441$-$00 & 197.5 & $-$28.9 & 0.4 & 4 & 2 & 0 & 1.0 & 2FGL J0442.7$-$0017 & \nodata & PKS 0440$-$00 &\\
J0448+11 & 187.5 & $-$21.1 & 0.4 & 2 & 2 & 0 & 1.0 & 2FGL J0448.9+1121 & \nodata & PKS 0446+11 &\\
J0456$-$23 & 223.8 & $-$35.1 & 0.2 & 34 & 11 & 14 & 0.5 & 2FGL J0457.0$-$2325 & \nodata & PKS 0454$-$234 &\\
J0503$-$01 & 201.6 & $-$24.7 & 0.6 & 2 & 1 & 0 & 1.4 & 2FGL J0501.2$-$0155 & 4396 & PKS 0458$-$02 &\\
J0504+04 & 195.6 & $-$21.2 & 0.3 & 5 & 4 & 0 & 0.7 & 2FGL J0505.5+0501 & \nodata & PKS 0502+049 &\\
J0517+45 & 163.1 & 4.4 & 0.6 & 2 & 1 & 0 & 1.4 & 2FGL J0517.0+4532 & \nodata & 4C +45.08 &\\
J0520$-$36 & 240.4 & $-$33.2 & 0.6 & 3 & 1 & 0 & 1.4 & 2FGL J0523.0$-$3628 & \nodata & PKS 0521$-$36 &\\
J0525+16 & 188.5 & $-$10.9 & 0.8 & 1 & 0 & 0 & 1.8 & \nodata & \nodata & \nodata &\\
J0532$-$48 & 255.0 & $-$32.6 & 0.2 & 9 & 7 & 0 & 0.6 & 2FGL J0532.0$-$4826 & \nodata & PMN J0531$-$4827 &\\
J0532+21 & 184.7 & $-$6.3 & 0.6 & 3 & 1 & 0 & 1.4 & 2FGL J0534.5+2201 & \nodata & PSR J0534+2200 &\\
J0533+07 & 197.2 & $-$13.8 & 0.3 & 4 & 3 & 0 & 0.8 & 2FGL J0532.7+0733 & \nodata & OG 050 &\\
J0536$-$44 & 250.2 & $-$31.6 & 0.2 & 16 & 10 & 1 & 0.5 & 2FGL J0538.8$-$4405 & \nodata & PKS 0537$-$441 &\\
J0537$-$54 & 262.8 & $-$32.4 & 0.6 & 1 & 1 & 0 & 1.4 & 2FGL J0540.4$-$5415 & \nodata & PKS 0539$-$543 &\\
J0537+13 & 191.9 & $-$9.3 & 0.4 & 5 & 2 & 0 & 1.0 & Fermi J0539+1432 & 3999 & TXS 0536+145 &\\
J0539$-$34 & 238.9 & $-$28.9 & 0.8 & 1 & 0 & 0 & 1.8 & 2FGL J0536.2$-$3348 & \nodata & 1RXS J053629.4$-$334302 &\\
J0604$-$70 & 280.6 & $-$29.4 & 0.6 & 1 & 1 & 0 & 1.4 & 2FGL J0601.1$-$7037 & \nodata & PKS 0601$-$70 &\\
J0622+33 & 179.7 & 9.2 & 0.2 & 16 & 13 & 0 & 0.5 & 2FGL J0622.9+3326 & \nodata & B2 0619+33 &\\
J0628$-$20 & 228.8 & $-$13.9 & 0.3 & 3 & 3 & 0 & 0.8 & 2FGL J0629.3$-$2001 & \nodata & PKS 0627$-$199 &\\
J0640+05 & 206.5 & 0.2 & 0.6 & 1 & 1 & 0 & 1.4 & Fermi J0639+0548 & 4224 & Nova Mon 2012 &\\
J0646$-$30 & 239.9 & $-$14.1 & 0.8 & 1 & 0 & 0 & 1.8 & Fermi J0648$-$3044 & 3878 & PKS 0646$-$306 &\\
J0652+45 & 171.2 & 18.9 & 0.3 & 4 & 3 & 0 & 0.8 & 2FGL J0654.2+4514 & \nodata & B3 0650+453 &\\
J0703$-$46 & 257.1 & $-$17.5 & 0.6 & 1 & 1 & 0 & 1.4 & 2FGL J0701.7$-$4630 & \nodata & PKS 0700$-$465 &\\
J0712+19 & 197.6 & 13.2 & 0.4 & 2 & 2 & 0 & 1.0 & 2FGL J0714.0+1933 & \nodata & MG2 J071354+1934 &\\
J0717+71 & 144.2 & 27.7 & 0.2 & 14 & 8 & 3 & 0.6 & 2FGL J0721.9+7120 & \nodata & S5 0716+71 &\\
J0720+33 & 185.2 & 20.1 & 0.3 & 4 & 3 & 0 & 0.8 & 2FGL J0719.3+3306 & \nodata & B2 0716+33 &\\
J0724+14 & 203.8 & 13.6 & 0.2 & 10 & 8 & 1 & 0.6 & 2FGL J0725.3+1426 & \nodata & 4C +14.23 &\\
J0728$-$11 & 227.6 & 2.8 & 0.4 & 4 & 0 & 0 & 0.9 & 2FGL J0730.2$-$1141 & \nodata & PKS 0727$-$11 &\\
J0730+36 & 181.9 & 23.3 & 0.6 & 1 & 1 & 0 & 1.4 & \nodata & \nodata & \nodata &\\
J0740+54 & 162.8 & 28.8 & 0.3 & 11 & 5 & 0 & 0.7 & 2FGL J0742.6+5442 & \nodata & GB6 J0742+5444 &\\
J0746+01 & 217.9 & 12.9 & 0.8 & 1 & 0 & 0 & 1.8 & 2FGL J0739.2+0138 & \nodata & PKS 0736+01 &\\
J0747+24 & 196.4 & 22.6 & 0.6 & 1 & 1 & 0 & 1.4 & \nodata & \nodata & \nodata &\\
J0759$-$56 & 270.0 & $-$13.7 & 0.6 & 1 & 1 & 0 & 1.4 & GRB080916C & \nodata & \nodata &\\
J0804+61 & 155.3 & 32.2 & 0.6 & 1 & 1 & 0 & 1.4 & 2FGL J0805.5+6145 & \nodata & TXS 0800+618 &\\
J0806+52 & 165.8 & 32.4 & 0.6 & 2 & 1 & 0 & 1.4 & \nodata & \nodata & \nodata &\\
J0808$-$08 & 229.4 & 13.0 & 0.2 & 6 & 6 & 0 & 0.6 & 2FGL J0808.2$-$0750 & \nodata & PKS 0805$-$07 &\\
J0811+02 & 220.0 & 18.9 & 0.6 & 1 & 1 & 0 & 1.4 & \nodata & \nodata & \nodata &\\
J0831+04 & 220.9 & 24.1 & 0.6 & 1 & 1 & 0 & 1.4 & 2FGL J0831.9+0429 & \nodata & PKS 0829+046 &\\
J0839+00 & 225.3 & 24.4 & 0.6 & 1 & 1 & 0 & 1.4 & 2FGL J0839.6+0059 & \nodata & PKS 0837+012 &\\
J0845+70 & 143.6 & 34.8 & 0.6 & 9 & 1 & 0 & 1.4 & 2FGL J0841.6+7052 & \nodata & 4C +71.07 &\\
J0849$-$12 & 238.6 & 19.4 & 0.3 & 3 & 3 & 0 & 0.8 & 2FGL J0850.2$-$1212 & \nodata & BZQ J0850$-$1213 &\\
J0849+50 & 168.3 & 39.1 & 0.2 & 7 & 6 & 0 & 0.6 & Fermi J0849+5108 & 3452 & SBS 0846+513 &\\
J0855+20 & 206.4 & 36.0 & 0.3 & 7 & 4 & 0 & 0.7 & 2FGL J0854.8+2005 & \nodata & OJ 287 &\\
J0905$-$35 & 259.9 & 7.5 & 0.6 & 1 & 1 & 0 & 1.4 & 2FGL J0904.8$-$3513 & \nodata & NVSS J090442$-$351423 &\\
J0907$-$02 & 232.7 & 28.5 & 0.8 & 1 & 0 & 0 & 1.8 & 2FGL J0909.7$-$0229 & \nodata & PKS 0907$-$023 &\\
J0908$-$50 & 271.0 & $-$1.6 & 0.6 & 1 & 1 & 0 & 1.4 & 2FGL J0910.4$-$5050 & \nodata & AT20G J091058$-$504807 &\\
J0908+60 & 155.0 & 40.0 & 0.6 & 1 & 1 & 0 & 1.4 & \nodata & \nodata & \nodata &\\
J0910+01 & 228.9 & 31.4 & 0.6 & 1 & 1 & 0 & 1.4 & 2FGL J0909.1+0121 & \nodata & PKS 0906+01 &\\
J0920+44 & 175.9 & 44.8 & 0.3 & 12 & 5 & 0 & 0.7 & 2FGL J0920.9+4441 & \nodata & S4 0917+44 &\\
J0924+28 & 199.0 & 44.5 & 0.6 & 1 & 1 & 0 & 1.4 & 2FGL J0924.0+2819 & \nodata & B2 0920+28 &\\
J0948+01 & 235.8 & 39.1 & 0.4 & 3 & 0 & 0 & 1.1 & 2FGL J0948.8+0020 & \nodata & PMN J0948+0022 &\\
J0955+65 & 145.7 & 42.6 & 0.6 & 1 & 1 & 0 & 1.4 & 2FGL J0958.6+6533 & \nodata & S4 0954+65 &\\
J0957$-$13 & 251.7 & 31.2 & 0.6 & 1 & 1 & 0 & 1.4 & 2FGL J0957.6$-$1350 & \nodata & PMN J0957$-$1350 &\\
J0957+24 & 206.9 & 50.9 & 0.6 & 1 & 1 & 0 & 1.4 & 2FGL J0956.9+2516 & \nodata & OK 290 &\\
J1013+24 & 208.4 & 54.5 & 0.6 & 1 & 1 & 0 & 1.4 & 2FGL J1012.6+2440 & \nodata & MG2 J101241+2439 &\\
J1015+04 & 236.9 & 46.8 & 0.6 & 1 & 1 & 0 & 1.4 & 2FGL J1016.0+0513 & \nodata & TXS 1013+054 &\\
J1023$-$31 & 269.6 & 21.2 & 0.6 & 1 & 1 & 0 & 1.4 & \nodata & \nodata & \nodata &\\
J1033+60 & 147.8 & 49.0 & 0.2 & 8 & 6 & 0 & 0.6 & 2FGL J1033.9+6050 & \nodata & S4 1030+61 &\\
J1038$-$53 & 283.6 & 4.7 & 0.3 & 4 & 3 & 0 & 0.8 & Fermi J1038$-$5314 & 3978 & PMN J1038$-$5311 &\\
J1040+06 & 240.9 & 52.6 & 0.6 & 1 & 1 & 0 & 1.4 & 2FGL J1040.7+0614 & \nodata & 4C +06.41 &\\
J1045+81 & 128.4 & 34.4 & 0.6 & 1 & 1 & 0 & 1.4 & 2FGL J1042.6+8053 & \nodata & S5 1039+81 &\\
J1046$-$29 & 272.6 & 26.1 & 0.6 & 3 & 1 & 0 & 1.4 & 2FGL J1045.5$-$2931 & \nodata & PKS B1043$-$291 &\\
J1051+04 & 245.7 & 53.6 & 0.6 & 1 & 1 & 0 & 1.4 & \nodata & \nodata & \nodata &\\
J1100+00 & 252.9 & 52.7 & 0.8 & 1 & 0 & 0 & 1.8 & 2FGL J1058.4+0133 & \nodata & 4C +01.28 &\\
J1102+37 & 180.8 & 64.9 & 0.6 & 1 & 1 & 0 & 1.4 & 2FGL J1104.4+3812 & \nodata & Mkn 421 &\\
J1110+34 & 187.1 & 67.1 & 0.8 & 1 & 0 & 0 & 1.8 & 2FGL J1112.4+3450 & \nodata & TXS 1109+350 &\\
J1120$-$05 & 265.7 & 50.7 & 0.3 & 5 & 3 & 0 & 0.8 & 2FGL J1121.5$-$0554 & \nodata & PKS 1118$-$05 &\\
J1124$-$64 & 293.8 & $-$3.1 & 0.8 & 1 & 0 & 0 & 1.8 & 1FGL J1122.9$-$6415 & \nodata & PMN J1123$-$6417 &\\
J1129$-$19 & 277.5 & 39.5 & 0.3 & 6 & 5 & 0 & 0.7 & 2FGL J1126.6$-$1856 & 3207 & PKS 1124$-$186 &\\
J1136$-$06 & 271.8 & 52.1 & 0.8 & 1 & 0 & 0 & 1.8 & \nodata & \nodata & \nodata &\\
J1145+39 & 165.6 & 71.3 & 0.3 & 5 & 3 & 0 & 0.8 & 2FGL J1146.9+4000 & \nodata & S4 1144+40 &\\
J1152+49 & 145.4 & 64.8 & 0.3 & 8 & 5 & 0 & 0.7 & 2FGL J1153.2+4935 & \nodata & OM 484 &\\
J1200+29 & 197.9 & 78.6 & 0.2 & 9 & 6 & 0 & 0.6 & 2FGL J1159.5+2914 & \nodata & Ton 599 &\\
J1205+54 & 136.6 & 61.0 & 0.6 & 2 & 1 & 0 & 1.4 & 2FGL J1208.8+5441 & \nodata & TXS 1206+549 &\\
J1217+29 & 190.3 & 82.0 & 0.6 & 1 & 1 & 0 & 1.4 & 2FGL J1217.8+3006 & \nodata & 1ES 1215+303 &\\
J1224+21 & 255.1 & 81.6 & 0.1 & 111 & 27 & 82 & 0.5 & 2FGL J1224.9+2122 & \nodata & 4C +21.35 &\\
J1229+02 & 290.3 & 64.5 & 0.2 & 34 & 10 & 14 & 0.6 & 2FGL J1229.1+0202 & \nodata & 3C 273 &\\
J1240+04 & 295.9 & 67.1 & 0.4 & 4 & 2 & 0 & 1.0 & 2FGL J1239.5+0443 & \nodata & MG1 J123931+0443 &\\
J1246$-$25 & 301.5 & 37.0 & 0.2 & 8 & 6 & 0 & 0.6 & 2FGL J1246.7$-$2546 & \nodata & PKS 1244$-$255 &\\
J1257$-$05 & 305.5 & 57.0 & 0.2 & 32 & 10 & 10 & 0.5 & 2FGL J1256.1$-$0547 & \nodata & 3C 279 &\\
J1300$-$22 & 305.6 & 40.1 & 0.6 & 1 & 1 & 0 & 1.4 & 2FGL J1258.8$-$2223 & \nodata & PKS 1256$-$220 &\\
J1303$-$64 & 304.3 & $-$2.1 & 0.3 & 5 & 0 & 0 & 0.9 & Fermi J1302$-$6350 & 3085 & PSRB 1259$-$63 &\\
J1312+55 & 116.7 & 61.6 & 0.6 & 1 & 1 & 0 & 1.4 & \nodata & \nodata & \nodata &\\
J1313+32 & 83.2 & 82.9 & 0.3 & 3 & 3 & 0 & 0.8 & 2FGL J1310.6+3222 & \nodata & OP 313 &\\
J1313+48 & 113.4 & 67.9 & 0.2 & 10 & 7 & 0 & 0.6 & 2FGL J1312.8+4828 & \nodata & GB 1310+487 &\\
J1318$-$33 & 309.5 & 28.9 & 0.6 & 2 & 1 & 0 & 1.4 & 2FGL J1315.9$-$3339 & \nodata & PKS 1313$-$333 &\\
J1329$-$70 & 306.1 & $-$7.5 & 0.4 & 3 & 2 & 0 & 1.0 & 2FGL J1330.1$-$7002 & \nodata & PKS 1326$-$697 &\\
J1331$-$55 & 308.6 & 6.6 & 0.8 & 1 & 0 & 0 & 1.8 & 2FGL J1329.2$-$5608 & \nodata & PMN J1329$-$5608 &\\
J1332$-$04 & 321.7 & 56.4 & 0.2 & 12 & 7 & 1 & 0.6 & 2FGL J1332.0$-$0508 & \nodata & PKS 1329$-$049 &\\
J1333+04 & 328.9 & 65.4 & 0.6 & 1 & 1 & 0 & 1.4 & \nodata & \nodata & \nodata &\\
J1346+44 & 94.7 & 69.1 & 0.2 & 13 & 8 & 0 & 0.6 & 2FGL J1345.4+4453 & \nodata & B3 1343+451 &\\
J1348$-$29 & 317.6 & 31.6 & 0.6 & 1 & 1 & 0 & 1.4 & 2FGL J1351.3$-$2909 & \nodata & PKS 1348$-$289 &\\
J1351$-$11 & 325.6 & 49.0 & 0.3 & 4 & 4 & 0 & 0.7 & Fermi J1349$-$1132 & 3788 & PKS 1346$-$112 &\\
J1353+30 & 49.4 & 76.1 & 0.6 & 1 & 1 & 0 & 1.4 & 2FGL J1350.8+3035 & \nodata & B2 1348+30B &\\
J1419+35 & 63.8 & 69.3 & 0.3 & 4 & 4 & 0 & 0.7 & \nodata & \nodata & \nodata &\\
J1428$-$41 & 321.6 & 17.5 & 0.2 & 14 & 6 & 4 & 0.6 & 2FGL J1428.0$-$4206 & \nodata & PKS B1424$-$418 &\\
J1459$-$35 & 330.2 & 20.4 & 0.2 & 6 & 6 & 0 & 0.6 & 2FGL J1457.4$-$3540 & \nodata & PKS 1454$-$354 &\\
J1505+10 & 11.5 & 54.4 & 0.1 & 73 & 26 & 40 & 0.3 & 2FGL J1504.3+1029 & \nodata & PKS 1502+106 &\\
J1513$-$09 & 351.4 & 40.0 & 0.1 & 110 & 19 & 76 & 0.5 & 2FGL J1512.8$-$0906 & \nodata & PKS 1510$-$08 &\\
J1517$-$31 & 335.9 & 21.8 & 0.8 & 1 & 0 & 0 & 1.8 & 2FGL J1513.6$-$3233 & \nodata & PKS 1510$-$324 &\\
J1522+31 & 50.0 & 56.8 & 0.2 & 13 & 7 & 1 & 0.6 & 2FGL J1522.1+3144 & \nodata & B2 1520+31 &\\
J1533$-$13 & 352.2 & 33.5 & 0.2 & 9 & 9 & 0 & 0.5 & Fermi J1532$-$1319 & 3579 & TXS 1530$-$131 &\\
J1554+13 & 24.3 & 45.2 & 0.4 & 3 & 2 & 0 & 1.0 & 2FGL J1553.5+1255 & \nodata & PKS 1551+130 &\\
J1626$-$25 & 352.0 & 16.0 & 0.3 & 6 & 4 & 0 & 0.7 & 2FGL J1625.7$-$2526 & \nodata & PKS 1622$-$253 &\\
J1636+38 & 61.3 & 42.1 & 0.2 & 24 & 6 & 6 & 0.6 & 2FGL J1635.2+3810 & \nodata & 4C +38.41 &\\
J1641+41 & 65.0 & 41.2 & 0.6 & 2 & 1 & 1 & 1.4 & 2FGL J1642.9+3949 & \nodata & 3C 345 &\\
J1641+47 & 73.8 & 41.3 & 0.6 & 1 & 1 & 0 & 1.4 & 2FGL J1637.7+4714 & \nodata & 4C +47.44 &\\
J1656+48 & 74.6 & 38.6 & 0.6 & 1 & 1 & 0 & 1.4 & 2FGL J1657.9+4809 & \nodata & 4C +48.41 &\\
J1700+68 & 99.7 & 35.2 & 0.3 & 5 & 3 & 0 & 0.8 & 2FGL J1700.2+6831 & \nodata & TXS 1700+685 &\\
J1704$-$62 & 328.1 & $-$12.7 & 0.8 & 1 & 0 & 0 & 1.8 & 2FGL J1703.2$-$6217 & \nodata & CGRaBS J1703$-$6212 &\\
J1707+77 & 109.3 & 32.1 & 0.8 & 1 & 0 & 0 & 1.8 & \nodata & \nodata & \nodata &\\
J1710+43 & 68.5 & 36.0 & 0.2 & 11 & 10 & 0 & 0.5 & 2FGL J1709.7+4319 & \nodata & B3 1708+433 &\\
J1715$-$33 & 352.5 & 2.8 & 0.4 & 2 & 2 & 0 & 1.0 & 2FGL J1717.7$-$3342 & \nodata & TXS 1714$-$336 &\\
J1718$-$52 & 337.5 & $-$8.5 & 0.4 & 2 & 2 & 0 & 1.0 & Fermi J1717$-$5156 & 4023 & PMN J1717$-$5155 &\\
J1719+18 & 39.8 & 28.1 & 0.4 & 2 & 2 & 0 & 1.0 & 2FGL J1719.3+1744 & \nodata & PKS 1717+177 &\\
J1732$-$12 & 12.1 & 10.9 & 0.3 & 3 & 3 & 0 & 0.8 & 2FGL J1733.1$-$1307 & \nodata & PKS 1730$-$13 &\\
J1735+39 & 64.3 & 30.7 & 0.3 & 4 & 4 & 0 & 0.7 & 2FGL J1734.3+3858 & \nodata & B2 1732+38A &\\
J1740+27 & 51.5 & 26.9 & 0.6 & 1 & 1 & 0 & 1.4 & GRB090902B & \nodata & \nodata &\\
J1741+50 & 77.1 & 31.5 & 0.6 & 1 & 1 & 0 & 1.4 & 2FGL J1739.5+4955 & \nodata & S4 1738+49 &\\
J1743+52 & 79.5 & 31.4 & 0.4 & 2 & 2 & 0 & 1.0 & 2FGL J1740.2+5212 & \nodata & 4C +51.37 &\\
J1746+70 & 100.7 & 30.9 & 0.3 & 4 & 3 & 0 & 0.8 & 2FGL J1748.8+7006 & 3171 & S4 1749+70 &\\
J1751+09 & 34.6 & 17.6 & 0.6 & 1 & 1 & 0 & 1.4 & 2FGL J1751.5+0938 & \nodata & OT 081 &\\
J1752$-$32 & 357.3 & $-$3.3 & 0.6 & 1 & 1 & 0 & 1.4 & Fermi J1750$-$3243 & 4284 & Nova Sco 2012 &\\
J1753+32 & 58.0 & 25.7 & 0.4 & 2 & 2 & 0 & 1.0 & 2FGL J1754.3+3212 & \nodata & RX J1754.1+3212 &\\
J1759$-$48 & 344.0 & $-$12.2 & 0.6 & 1 & 1 & 0 & 1.4 & 2FGL J1759.2$-$4819 & \nodata & PMN J1758$-$4820 &\\
J1803$-$39 & 352.9 & $-$8.4 & 0.2 & 8 & 6 & 0 & 0.6 & 2FGL J1802.6$-$3940 & \nodata & PMN J1802$-$3940 &\\
J1807+78 & 110.1 & 28.7 & 0.3 & 7 & 5 & 0 & 0.7 & 2FGL J1800.5+7829 & \nodata & S5 1803+784 &\\
J1825+56 & 85.8 & 25.9 & 0.4 & 2 & 2 & 0 & 1.0 & 2FGL J1824.0+5650 & \nodata & 4C +56.27 &\\
J1827$-$52 & 342.6 & $-$17.8 & 0.6 & 1 & 1 & 0 & 1.4 & 2FGL J1825.1$-$5231 & \nodata & PKS 1821$-$525 &\\
J1834$-$21 & 12.3 & $-$5.9 & 0.2 & 10 & 10 & 0 & 0.5 & 2FGL J1833.6$-$2104 & \nodata & PKS 1830$-$211 &\\
J1850+32 & 62.2 & 14.3 & 0.2 & 7 & 6 & 0 & 0.6 & 2FGL J1848.5+3216 & \nodata & B2 1846+32A &\\
J1853+48 & 78.3 & 19.7 & 0.6 & 1 & 1 & 0 & 1.4 & 2FGL J1852.5+4856 & \nodata & S4 1851+48 &\\
J1853+67 & 97.6 & 24.7 & 0.2 & 13 & 11 & 0 & 0.5 & 2FGL J1849.4+6706 & \nodata & S4 1849+67 &\\
J1909$-$80 & 313.9 & $-$27.5 & 0.6 & 1 & 1 & 0 & 1.4 & \nodata & \nodata & \nodata &\\
J1910$-$20 & 16.8 & $-$13.0 & 0.5 & 2 & 0 & 0 & 1.3 & 2FGL J1911.1$-$2005 & \nodata & PKS B1908$-$201 &\\
J1914$-$35 & 1.8 & $-$19.9 & 0.6 & 1 & 1 & 0 & 1.4 & Fermi J1913$-$3630 & 2966 & PMN J1913$-$3630 &\\
J1924$-$21 & 17.1 & $-$16.4 & 0.3 & 5 & 3 & 0 & 0.8 & 2FGL J1923.5$-$2105 & \nodata & TXS 1920$-$211 &\\
J1956$-$38 & 1.9 & $-$28.5 & 0.6 & 1 & 1 & 0 & 1.4 & 2FGL J1958.2$-$3848 & \nodata & PKS 1954$-$388 &\\
J1956$-$42 & 356.9 & $-$29.6 & 0.6 & 2 & 1 & 0 & 1.4 & 2FGL J1959.1$-$4245 & \nodata & PMN J1959$-$4246 &\\
J2001+44 & 79.3 & 7.3 & 0.6 & 1 & 1 & 0 & 1.4 & 2FGL J2001.1+4352 & \nodata & MAGIC J2001+435 &\\
J2012+37 & 74.6 & 1.7 & 0.6 & 3 & 1 & 0 & 1.4 & 2FGL J2015.6+3709 & \nodata & MG2 J201534+3710 &\\
J2023+33 & 73.0 & $-$2.1 & 0.6 & 1 & 1 & 0 & 1.4 & 2FGL J2025.1+3341 & \nodata & B2 2023+33 &\\
J2026$-$07 & 37.0 & $-$24.6 & 0.2 & 15 & 8 & 0 & 0.6 & 2FGL J2025.6$-$0736 & \nodata & PKS 2023$-$07 &\\
J2034+40 & 80.0 & 0.4 & 0.4 & 3 & 0 & 0 & 1.1 & 2FGL J2032.1+4049 & \nodata & Cyg X$-$3 &\\
J2036+11 & 55.7 & $-$17.3 & 0.6 & 1 & 1 & 0 & 1.4 & 2FGL J2035.4+1058 & \nodata & PKS 2032+107 &\\
J2056$-$47 & 352.4 & $-$40.4 & 0.3 & 6 & 4 & 0 & 0.7 & 2FGL J2056.2$-$4715 & \nodata & PKS 2052$-$47 &\\
J2103+45 & 87.1 & $-$0.7 & 0.4 & 2 & 2 & 0 & 1.0 & 2FGL J2102.2+4546 & \nodata & V407 Cyg &\\
J2122$-$46 & 353.8 & $-$44.9 & 0.8 & 1 & 0 & 0 & 1.8 & 2FGL J2125.0$-$4632 & 3808 & PKS 2123$-$463 &\\
J2134$-$01 & 52.3 & $-$36.5 & 0.6 & 1 & 1 & 0 & 1.4 & 2FGL J2133.8$-$0154 & 4333 & PKS 2131$-$021 &\\
J2144+17 & 72.2 & $-$26.2 & 0.8 & 1 & 0 & 0 & 1.8 & 2FGL J2143.5+1743 & \nodata & OX 169 &\\
J2148$-$75 & 315.9 & $-$36.7 & 0.3 & 10 & 5 & 0 & 0.7 & 2FGL J2147.4$-$7534 & \nodata & PKS 2142$-$75 &\\
J2154$-$30 & 17.1 & $-$51.3 & 0.3 & 5 & 0 & 0 & 0.9 & 2FGL J2151.5$-$3021 & \nodata & PKS 2149$-$306 &\\
J2155$-$83 & 307.9 & $-$31.4 & 0.6 & 2 & 1 & 0 & 1.4 & 2FGL J2201.9$-$8335 & \nodata & PKS 2155$-$83 &\\
J2159+31 & 85.1 & $-$18.5 & 0.6 & 4 & 1 & 0 & 1.4 & 2FGL J2157.4+3129 & \nodata & B2 2155+31 &\\
J2202+42 & 92.5 & $-$10.4 & 0.2 & 15 & 8 & 1 & 0.6 & 2FGL J2202.8+4216 & \nodata & BL Lacertae &\\
J2202+50 & 97.7 & $-$3.6 & 0.3 & 6 & 4 & 0 & 0.7 & Fermi J2202+5045 & 4182 & NRAO 676 &\\
J2209$-$53 & 339.8 & $-$50.2 & 0.6 & 1 & 1 & 0 & 1.4 & 2FGL J2208.1$-$5345 & \nodata & PKS 2204$-$54 &\\
J2214$-$26 & 24.5 & $-$55.1 & 0.6 & 1 & 1 & 0 & 1.4 & GRB090510 & \nodata & \nodata &\\
J2229$-$07 & 56.1 & $-$51.3 & 0.5 & 2 & 0 & 0 & 1.3 & 2FGL J2229.7$-$0832 & \nodata & PKS 2227$-$08 &\\
J2233+11 & 77.7 & $-$38.7 & 0.3 & 4 & 3 & 0 & 0.8 & 2FGL J2232.4+1143 & \nodata & CTA 102 &\\
J2237$-$14 & 48.3 & $-$56.3 & 0.3 & 5 & 5 & 0 & 0.7 & 2FGL J2236.5$-$1431 & \nodata & PKS 2233$-$148 &\\
J2238$-$39 & 0.4 & $-$60.0 & 0.6 & 1 & 1 & 0 & 1.4 & \nodata & \nodata & \nodata &\\
J2247$-$06 & 62.9 & $-$53.7 & 0.8 & 1 & 0 & 0 & 1.8 & \nodata & \nodata & \nodata &\\
J2252$-$27 & 24.5 & $-$63.6 & 0.3 & 7 & 3 & 0 & 0.8 & 2FGL J2250.8$-$2808 & \nodata & PMN J2250$-$2806 &\\
J2254+16 & 86.2 & $-$38.2 & 0.1 & 168 & 137 & 123 & 0.5 & 2FGL J2253.9+1609 & \nodata & 3C 454.3 &\\
J2311+34 & 100.5 & $-$24.1 & 0.3 & 9 & 5 & 0 & 0.7 & 2FGL J2311.0+3425 & \nodata & B2 2308+34 &\\
J2321+32 & 101.7 & $-$27.1 & 0.4 & 2 & 2 & 0 & 1.0 & 2FGL J2322.2+3206 & \nodata & B2 2319+31 &\\
J2322$-$03 & 76.8 & $-$58.6 & 0.6 & 1 & 1 & 0 & 1.4 & 2FGL J2323.6$-$0316 & \nodata & PKS 2320$-$035 &\\
J2326+40 & 105.8 & $-$19.6 & 0.6 & 1 & 1 & 0 & 1.4 & 2FGL J2325.3+3957 & \nodata & B3 2322+396 &\\
J2328$-$49 & 332.4 & $-$62.3 & 0.3 & 27 & 3 & 2 & 0.8 & 2FGL J2329.2$-$4956 & \nodata & PKS 2326$-$502 &\\
J2329$-$21 & 45.3 & $-$70.5 & 0.6 & 1 & 1 & 0 & 1.4 & 2FGL J2330.9$-$2144 & \nodata & PMN J2331$-$2148 &\\
J2330+09 & 91.8 & $-$48.5 & 0.6 & 3 & 1 & 0 & 1.4 & 2FGL J2327.5+0940 & \nodata & PKS 2325+093 &\\
J2332$-$66 & 314.8 & $-$49.0 & 0.6 & 1 & 1 & 0 & 1.4 & GRB090926A & \nodata & \nodata &\\
J2333$-$40 & 348.4 & $-$69.1 & 0.4 & 2 & 2 & 0 & 1.0 & 2FGL J2336.3$-$4111 & \nodata & PKS 2333$-$415 &\\
J2345$-$15 & 65.8 & $-$71.0 & 0.1 & 20 & 17 & 1 & 0.4 & 2FGL J2345.0$-$1553 & \nodata & PMN J2345$-$1555 &\\

\enddata 
\label{tab:cat}
\end{deluxetable} 
\end{landscape}
\end{document}